\documentclass[11pt]{article}

\usepackage{latexsym,amsmath,epsfig}
     
\DeclareFontFamily{OT1}{rsfs10}{}
\DeclareFontShape{OT1}{rsfs10}{m}{n}{ <-> rsfs10 }{}
\DeclareMathAlphabet{\mathscript}{OT1}{rsfs10}{m}{n}

\numberwithin{equation}{section}

\newcommand{\be}{\begin{equation}}
\newcommand{\ee}{\end{equation}}

\newcommand{\bea}{\begin{eqnarray}}
\newcommand{\eea}{\end{eqnarray}}
\newcommand{\ba}{\begin{array}}
\newcommand{\ea}{\end{array}}

\newcommand{\ns}{\normalsize}
\newcommand{\pt}{\partial}

\def\a{\alpha}
\def\b{\beta}
\def\g{\gamma}
\def\c{\chi}
\def\d{\delta}
\def\e{\epsilon}

\def\f{\phi}
\def\z{\psi}

\def\k{\kappa}

\def\m{\mu}
\def\n{\nu}

\def\p{\pi}

\def\r{\rho}
\def\s{\sigma}

\def\x{\xi}
\def\z{\zeta}

\def\F{\Phi}
\def\G{\Gamma}

\begin{document}

\begin{titlepage}

\title{
\hfill{\ns SUSX-TH/02-018\\}
\hfill{\ns hep-th/0208219\\[2cm]}
{\LARGE Kinky Brane Worlds }\\[1cm]}
\setcounter{footnote}{0}
\author{{\ns\large
 Nuno D. Antunes\footnote{email: mppg5@pact.cpes.susx.ac.uk}~,
 Edmund J.~Copeland\footnote{email: e.j.copeland@sussex.ac.uk}~,
\setcounter{footnote}{3}
 Mark Hindmarsh\footnote{email: m.b.hindmarsh@sussex.ac.uk}~~and
 Andr\'e Lukas\footnote{email: a.lukas@sussex.ac.uk}} \\[0.8em]
      {\ns Centre for Theoretical Physics,
     University of Sussex}\\[-0.2em]
      {\ns Falmer, Brighton BN1 9QJ, United Kingdom}}


\maketitle

\vspace{1cm}

\begin{abstract}

We present a toy model for five-dimensional heterotic M-theory where
bulk three-branes, originating in 11 dimensions from M five-branes,
are modelled as kink solutions of a bulk scalar field theory. It is
shown that the vacua of this defect model correspond to a class of
topologically distinct M-theory compactifications. Topology change can
then be analysed by studying the time evolution of the defect model.
In the context of a four-dimensional effective theory, we study in
detail the simplest such process, that is the time evolution of a kink
and its collision with a boundary. We find that the kink is
generically absorbed by the boundary thereby changing the boundary
charge. This opens up the possibility of exploring the relation
between more complicated defect configurations and the topology of
brane-world models.

\end{abstract}

\thispagestyle{empty}

\end{titlepage}


\section{Introduction}
     
The single most important problem in trying to make contact
between string-/M-theory and low-energy physics is probably 
the large number of degenerate and topologically distinct vacua of the
theory. It is usually stated that non-perturbative effects
will eventually lift most of this degeneracy. However, despite
the advances over recent years in understanding non-perturbative
string- and M-theory there is very little indication
of progress in this direction. In fact, with the advent of
M-theory and concepts such as branes and brane-world theories
new classes of vacua have been constructed and, as a consequence,
the degeneracy problem has perhaps grown even more serious.
It seems worthwhile, therefore, to ask whether the cosmological
evolution rather than inherent non-perturbative effects of the
theory may play a prominent role in selecting the vacuum state.
Indeed, it is known that the degeneracy of some vacua (particularly
among those with a large number of supersymmetries) will not
be lifted non-perturbatively, suggesting cosmology will have
some role to play.

\vspace{0.4cm}

The first task to tackle, in this context, is the formulation
of a workable theory capable of describing a number of
topologically different vacua and transitions among them.
As a second step, one will have to analyse the cosmological
evolution of this theory. It is precisely these two problems
which will be the main topic of the present paper. 

The class of vacua we will use in our approach is provided by
compactification of heterotic M-theory~\cite{Horava:1996ma} on
Calabi-Yau three
folds~\cite{Witten:1996mz,Horava:1996vs,Lukas:1997fg,Lukas:1998hk}
resulting in five-dimensional brane-world
theories~\cite{Lukas:1998yy,Ellis:1998dh,Lukas:1998tt}.  These
theories are defined on a space-time with two four-dimensional
boundaries corresponding to the fixed planes of the orbifold $S^1/Z_2$
and, in addition, may contain bulk three-branes which originate from M
five-branes wrapping two-cycles in the Calabi-Yau
space~\cite{Witten:1996mz,Lukas:1998hk}.  The associated effective
actions are five-dimensional gauged $N=1$ supergravity theories in the
bulk coupled to four-dimensional $N=1$ theories residing on the two
boundaries and the three-branes. The prospects for particle-physics
model building within this class of compactifications is quite
promising and a number of models with attractive particle-physics
properties on the "observable" boundary have been
constructed~\cite{Andreas:1999ei}--\cite{Donagi:2000zs}.  The simplest
way to characterise topologically different compactifications from the
viewpoint of the five-dimensional effective theories is by using the
charges $\a_1$ and $\a_2$ on the boundaries and the three-brane charge
$\a_3$. These charges are not independent but must satisfy the
cohomology constraint $\a_1+\a_2+\a_3=0$ which follows from anomaly
cancellation. Two five-dimensional effective theories with different
sets of charges $(\a_1,\a_2,\a_3)$ originate from topologically
distinct compactifications.  A transition between two such theories
may occur through a small-instanton
transition~\cite{Witten:1996gx,Ganor:1996mu} when a three-brane
collides with one of the boundaries. The three-brane can then be
"absorbed" by the boundary and, correspondingly, the boundary charge
is changed by the amount carried by the incoming three-brane.  This
change in the boundary charge indicates a more dramatic transition in
the boundary theory. For example, the gauge group and the amount of
chiral matter~\cite{Ovrut:2000qi} may be altered as a consequence of the
internal topology change.

\vspace{0.4cm}

The goal of this paper is to find a five-dimensional (toy) model which
provides a unified description for the above class of topologically
distinct vacua, in the simplest setting, and allows for transitions
between them. While, for simplicity, we will assume that the topology
of space-time both in the internal Calabi-Yau space and in the
orbifold direction remains unchanged we will allow for transitions
corresponding to a topology change in the internal gauge-field
instantons on the boundaries and a change in the number and charges of
three-branes. Our basic method will be, starting with five-dimensional
heterotic M-theory in its simplest form, to model the three-branes as
topological defects~\cite{DeWolfe:1999cp} (kinks) of a new bulk scalar
field $\c$. We do not claim, of course, that this model provides the
correct definition of M-theory in these backgrounds.  However, we do
show that the defect model in the background of its various vacuum
states reproduces the five-dimensional M-theory effective actions with
different charges $(\a_1,\a_2,\a_3)$, corresponding to topologically
distinct M-theory compactifications.

\vspace{0.4cm}

Time-evolution of the defect model and the scalar $\c$
in particular then allows for a transition between these topologically
distinct configurations. We will study in detail the simplest such
transition, namely the collision of a three-brane kink with one
of the boundaries. This will be done by calculating the four-dimensional
effective action for the defect model in the background of such a kink.
As we will see from this four-dimensional action, the
collision  process indeed generically leads to an absorption of
the kink and a change in the boundary charge by the amount carried
by the kink. Hence, we have established the existence of one of
the elementary topology-changing processes in our defect model.
This opens up the possibility, subject of ongoing research, that a
study of more complicated configurations, such as brane-networks,
will provide insight into topological properties of brane-world models. 

\vspace{0.4cm}

The plan of the paper is as follows. In the next section, we will introduce
the five-dimensional effective actions from heterotic M-theory, in their
simplest form. For later reference, we will also review the associated
four-dimensional effective theories. Section 3 then presents our defect
model and explains how, precisely, it is related to the M-theory actions.
In Section 4, we will compute the four-dimensional effective action
for the defect model in the background of a kink and Section 5 presents
the resulting evolution equations. Section 6 is devoted to a detailed
study of the kink evolution and its collision with a boundary, based
on these equations. A conclusion and outlook is presented in Section 7.

\section{Effective actions from heterotic M-theory}

To set the scene, we will now describe the five-dimensional brane world
theories for which we would like to find a smooth defect-model.
These brane-world theories can be viewed as a minimal version
of five-dimensional heterotic M-theory~\cite{Lukas:1998yy}. For later
purposes, it will also be useful to review the four-dimensional
effective action associated to these brane-world theories.

\vspace{0.4cm}

Coordinates for the five-dimensional space $M_5$ are denoted by $x^\a$
where $\a ,\b,\dots = 0,1,2,3,5$. We also introduce four-dimensional
indices $\m ,\n ,\dots = 0,1,2,3$. The coordinate $y\equiv x^5$ is
compactified on an orbi-circle $S^1/Z_2$ in the usual way, that is, by
first compactifying $y$ on a circle with radius $\r$ and then dividing
by the $Z_2$ orbifold action $y\rightarrow -y$. Taking the
$y$--coordinate in the range $y\in [-\p\r ,\p\r ]$ with the endpoints
being identified the two resulting four-dimensional fixed planes
(boundaries), denoted by $M_4^1$ and $M_4^2$, are located at $y=0$ and
$y=\p\r$, respectively. Such a geometry is obtained by compactifying
11-dimensional heterotic M-theory on a Calabi-Yau space. If the
five-branes present in the 11-dimensional theory are included in this
compactification they lead, upon wrapping a two-cycle in the
Calabi-Yau space, to bulk three-branes in the five-dimensional
brane-world theories. For simplicity, we will consider a single such
three-brane whose world-volume we denote by $M_4^3$. We also need to
include the $Z_2$ mirror of this three-brane with world-volume
$\tilde{M}_4^3$. Three-brane world-volume coordinates are denoted by
$\s^\m$. In the minimal version of the model the bulk fields consist
of the metric and the dilaton $\F$ while the three-brane world-volume
fields are simply the embedding coordinates $X^\a =X^\a (\s^\m )$. The
effective action for these fields is then given by~\cite{Brandle:2001ts}
\begin{eqnarray}
 S_5 &=& -\frac{1}{2\k_5^2}\left\{\int_{M_5}\sqrt{-g}\left[
         \frac{1}{2}R+\frac{1}{4}\partial_\a\F\partial^\a\F
         +\frac{1}{3}\a^2 e^{-2\F}\right]\right. \nonumber \\
     &&\qquad +\int_{M_4^1}\sqrt{-g}\; 2\a_1 e^{-\F} 
       +\int_{M_4^2}\sqrt{-g}\; 2\a_2 e^{-\F} \nonumber \\
     &&\qquad \left. +\int_{M_4^3\cup\tilde{M}_4^3}\sqrt{-\g}\;
       |\a_3| e^{-\F}\right\}\; . \label{S5}
\end{eqnarray}
Note, that the dilaton $\F$ measures the size of the internal Calabi-Yau
space which is, more precisely, given by $v e^\F$, where $v$ is
a fixed reference volume. It relates the five-dimensional Newton
constant $\k_5$ to its 11-dimensional counterpart $\k$ via
\begin{equation}
 \k_5^2 = \frac{\k^2}{v}\; .
\end{equation}
Further, $\a_i$, where $i=1,2,3$ are the charges
on the orbifold planes and the three-brane, respectively. They are
quantised and can be written as integer multiples
\begin{equation}
 \a_i = \s\b_i\; ,\qquad \b_i\in{\bf Z}
\end{equation}
of the unit charge $\s$ defined by
\begin{equation}
\s = \frac{\e_0}{\p\r}\; ,\qquad
\e_0 = \left(\frac{\k}{4\p}\right)^{2/3}\frac{2\p^2\r}{v^{2/3}}\; .\label{e0}
\end{equation}
These charges satisfy the important cohomology condition
\begin{equation}
 \sum_{i=1}^3\a_i = 0 \label{cohomology}
\end{equation}
which follows from anomaly cancellation in the 11-dimensional theory.
The quantity $\a$ which appears in the above bulk potential is
a sum of step-functions given by
\begin{equation}
 \a = \a_1\theta (M_4^1) + \a_2\theta (M_4^2) + \a_3\left(\theta (M_4^3)+
      \theta (\tilde{M}_4^3)\right) \; . \label{alpha}
\end{equation}
Finally, the induced metric $\g_{\m\n}$ on the three-brane world-volume
is defined as the pull-back
\begin{equation}
 \g_{\m\n} = \partial_\m X^\a\partial_\n X^\b g_{\a\b}
\end{equation}
of the space-time metric.

For positive three-brane charge, $\a_3 > 0$, the above action can
be embedded into a five-dimensional $N=1$ bulk supergravity theory
coupled to four-dimensional $N=1$ theories on the boundaries and
the branes. The details of this supergravity theory have been
worked out in Ref.~\cite{Brandle:2001ts}. In the case of an anti-three-brane,
that is for $\a_3 < 0$, while bulk supersymmetry is preserved everywhere
locally, it is broken globally. Technically, this happens because
the chirality of the four-dimensional supersymmetry preserved on
the three-brane is opposite to the one on the orbifold fixed planes.
Such non-supersymmetric heterotic models containing anti-branes
have not been studied in much detail, so far. We have included this
possibility here because it will naturally arise later in our
discussion of the defect model. The generalisation
to include more than one three-brane is straightforward. It simply
amounts to replacing the Nambu-Goto type three-brane action 
in the third line of~\eqref{S5} by a sum over such actions (with
generally different three-brane charges) and modifying the cohomology
condition~\eqref{cohomology} and the definition of
$\a$, Eq.~\eqref{alpha}, accordingly.

Note that two actions of the type~\eqref{S5} but with different
sets of charges $\a_i$ correspond to topologically different
M-theory compactifications. Specifically, the charges $\a_1$
and $\a_2$ on the boundaries are related to gravitational and
gauge instanton numbers. If we keep the topology of the Calabi-Yau
space fixed, as discussed, different values of $\a_1$ and $\a_2$
indicate a different topology of the internal gauge bundles.
As a consequence, the values of $\a_1$, $\a_2$ are also correlated
with other properties of the boundary theories, such as the types
of gauge groups and the amount of chiral matter. Different values
of $\a_3$ imply different internal wrapping numbers for the five-branes
and, hence, clearly indicate different topologies.

\vspace{0.4cm}

For the case of a three-brane (rather than an anti-three-brane),
the action~\eqref{S5} has a BPS domain-wall
vacuum~\cite{Lukas:1998yy,Brandle:2001ts} given by
\begin{eqnarray}
 ds^2 &=& a_0^2hdx^\m dx^\n\eta_{\m\n}+b_0^2h^4dy^2 \label{BPS1}\\
 e^\F &=& b_0h^3 \label{BPS2}\\
 X^\m &=& \s^\m \\
 X^5 &=& Y = {\rm const}\label{BPS4}
\end{eqnarray}
Here the function $h=h(y)$ is defined by
\begin{equation}
  h(y)=-\frac{2}{3}\left\{ \begin{array}{lll}
      \a_{1} |y|+c_0 &\mbox{for}& 0\leq |y|\leq Y \\
      (\a_{1}+\a_{3})|y|-\a_{3}\, Y+c_0 &\mbox{for}&
      Y\leq |y|\leq \p\r
      \end{array}\right.\label{BPS5}
\end{equation}
and $a_0$, $b_0$ and $c_0$ are constants. Note that this solution
is not smooth across the three-brane reflecting the fact that the
three-brane as described by~\eqref{S5} is infinitely thin. Such a
static BPS solution does not exist for the anti-three-brane
since the sum of the tensions $\a_1+\a_2+|\a_3|$ does not vanish
for $\a_3<0$ by virtue of the cohomology condition~\eqref{cohomology}.
In fact, solutions which couple to an anti-three-brane will, in general,
be time-dependent.

For later reference, it will be useful to discuss the four-dimensional
effective action associated to the brane-world model~\eqref{S5} and
the above BPS vacuum. It is given
by~\cite{Derendinger:2001gy,Brandle:2001ts}
\begin{equation}
 S_4 = 
-\frac{1}{2\k_P^2}\int_{M_4}\sqrt{-g_4}\left[\frac{1}{2}R_4+\frac{1}{4}\pt_\m\f
     \pt^\m\f +\frac{3}{4}\pt_\m\b\pt^\m\b+\frac{q_3}{2}e^{\b -\f}
     \pt_\m z\pt^\m z\right]\; . \label{S4}
\end{equation}
The three scalar fields $\f$, $\b$ and $z$ have straightforward
interpretations in terms of the underlying higher-dimensional theories.
The field $\f$, as the zero mode of the five-dimensional scalar $\F$,
specifies the volume of the internal Calabi-Yau space averaged over
the orbifold. More precisely, this average volume is given by $ve^\f$.
The scalar $\b$, on the other hand, originates from the $(55)$-component
of the five-dimensional metric and measures the size $\p\r e^\b$ of
the orbifold. Finally, $z$ represents the position of the three-brane
and is normalised to be in the range $z\in [0,1]$ with the endpoints
corresponding to the two boundaries of five-dimensional space-time.
The four-dimensional Newton constant $\k_P$ is related to its five-dimensional
cousin by
\begin{equation}
 \k_P^2 = \frac{\k_5^2}{2\p\r}\; .
\end{equation}
Finally, the three-brane charge
\begin{equation}
 q_3=\p\r\a_3=\e_0\b_3\; , \qquad \b_3\in{\bf Z} \label{q3} 
\end{equation}
is quantised in units of $\e_0$ as defined in Eq.~\eqref{e0} and
is positive for the case under discussion.

As expected, the action~\eqref{S4} can be obtained from an $N=1$
supergravity theory by a suitable truncation.  The Kahler potential
for this supergravity theory has been first given in
Ref.~\cite{Derendinger:2001gy}. An important quantity which governs
the validity of the effective action~\eqref{S4} is the strong-coupling
expansion parameter
\begin{equation}
 \e = \e_0 e^{\b - \f}\; .
\end{equation}
It measures the relative size of string loop corrections to the
four-dimensional action~\eqref{S4} or, equivalently, the strength of
the warping in the orbifold direction from a five-dimensional
viewpoint. The effective action~\eqref{S4} is valid as long as $\e <1$
and can be expected to break down otherwise. Another reason for a
breakdown of the four- as well as the five-dimensional effective
theory is the five-brane approaching one of the boundaries, that is,
$z\rightarrow 0$ or $z\rightarrow 1$. In this case, the underlying
heterotic M-theory may undergo a small-instanton
transition~\cite{Witten:1996gx,Ganor:1996mu} which leads to the M
five-brane being converted into a gauge-field instanton (or, so called
gauge five-brane~\cite{Strominger:et}) on the boundary. In such a
process, properties of the boundary theory, such as the gauge group
and the amount of chiral matter, can change dramatically as a result
of the internal topology change~\cite{Ovrut:2000qi}. In our simple
five-dimensional model~\eqref{S5} such a modification of the boundary
theory is indicated by a change in the boundary charge $\a_1$ or
$\a_2$ by the amount of incoming five-brane charge. It is clear,
however, that the actions~\eqref{S5} or \eqref{S4} are not capable of
describing such a jump in the boundary charge in a dynamical way. In
fact, the four-dimensional action~\eqref{S4} does not retain any
memory of the presence of the boundaries as $z\rightarrow 0,1$. This
can also be seen from the moving-brane solutions to~\eqref{S4} found
in Ref.~\cite{Copeland:2001zp} and will be explained in more detail
later. As we will see, our defect model, to be presented in the next
section, will considerably improve on these points.


\section{Modelling heterotic brane-world theories}

We would now like to find a ``smooth'' model, replacing the
five-dimensional action~\eqref{S5}, where the three-brane is not
put in ``by hand'' but, rather, obtained as defect solution to
the theory. Such a model should have, as a solution,
a smooth version of the BPS domain wall~\eqref{BPS1}--\eqref{BPS5}.
Note, that we will not attempt to find a smooth description for the orbifold
fixed planes. Their nature, as part of the space-time geometry,
is entirely different from the one of the three-branes. In particular,
the fixed plane tensions $\a_1$, $\a_2$ can be negative
whereas the three-brane tension $|\a_3|$ is always positive. 

\begin{figure}\centering
\includegraphics[height=9cm,width=9cm, angle=0]{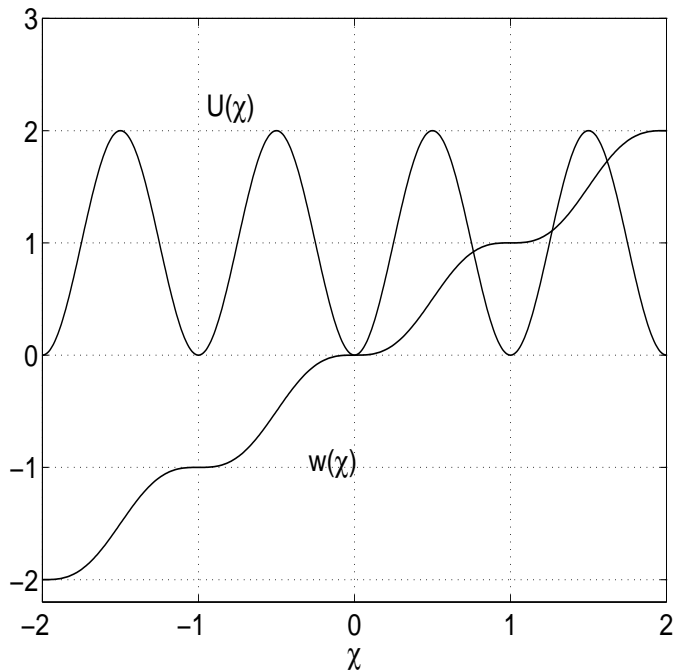}
        \caption{\emph{Shown is the typical shape of the superpotential
                       $w$ and the potential $U$ (in units of
                       $\sigma$) as a function of
                       the scalar field $\c$ (in units of $v$). }}
\label{fig1}
\end{figure}

Modelling co-dimension one objects such as our three-branes is usually
achieved using kink-solutions of scalar field theories~\cite{DeWolfe:1999cp}.
This is indeed what we will do here. We, therefore, supplement the
bulk field content of the five-dimensional theory by a second scalar
field $\c$. For this bulk scalar along with the dilaton $\F$ and the
five-dimensional metric, we propose the following action
\begin{eqnarray}
 \tilde{S}_5 &=& -\frac{1}{2\k_5^2}\left\{\int_{M_5}\sqrt{-g}\left[
                 \frac{1}{2}R+\frac{1}{4}\partial_\a\F\partial^\a\F
                 +\frac{1}{2}e^{-\F}\partial_\a\c\partial^\a\c
                 +V(\F ,\c )\right]\right.\nonumber \\
             &&\qquad\left. +\int_{M_4^1}\sqrt{-g}\; 2W
                            -\int_{M_4^2}\sqrt{-g}\; 2W\right\}\; .
 \label{S5t}
\end{eqnarray}
We require that the potential $V$ be obtained from
a ``superpotential'' $W$ following the general formula~\cite{Skenderis:1999mm}
\begin{equation}
 V = \frac{1}{2}G^{IJ}\partial_IW\partial_JW - \frac{2}{3}W^2\; ,
 \label{Vgen}
\end{equation}
where $G_{IJ}$ is the sigma-model metric and indices $I,J,\dots$
label the various scalar fields $\F^I$. For our specific action~\eqref{S5t},
we have two scalar fields $(\F^I)=(\F ,\c )$ and the sigma-model metric
is explicitly given by
\begin{equation}
 G = {\rm diag}\left(\frac{1}{2},e^{-\F}\right)\; .
\end{equation}
Further, we propose the following form for the superpotential
\begin{equation}
 W=e^{-\F}w(\c ) \label{W}
\end{equation}
where $w$ is an, as yet, unspecified function of $\c$. Using the
general expression~\eqref{Vgen} this results in a potential
\begin{equation}
 V = \frac{1}{3}e^{-2\F}w^2+\frac{1}{2}e^{-\F}U\; ,\qquad
 U = \left(\frac{dw}{d\c}\right)^2\; . \label{U}
\end{equation}
Note that, in~\eqref{S5t}, we have omitted the Nambu-Goto
type action for the three-brane corresponding to the third line of the
M-theory effective action~\eqref{S5}. The reason is, of course, that
we would like to recover the three-brane as a kink-solution of the new
scalar field $\c$. For this to work out, the potential $U$ has to have
a non-trivial vacuum structure. In fact, since the original
three-brane charge is an (arbitrary) integer multiple of a certain
unit, we need an infinite number of equally spaced minima. More
precisely, we require the potential $U$ satisfies the following properties~:
\begin{itemize}
 \item $U$ is periodic with period $v$, that is $U(\c +v)=U(\c)$
 \item $U$ has minima at $\c = \c_n = nv$ for all $n\in{\bf Z}$
 \item $U$ vanishes at the minima, that is $U(\c_n)=0$.
\end{itemize}
These requirements can be easily translated into conditions on the
function $w$ which determines the superpotential. Clearly, from the
second and third condition, the derivative of $w$ has to vanish at all
minima $\c_n=nv$ of $U$. The definition~\eqref{U} of $w$ in terms of
$U$ involves a sign ambiguity which allows one, using the first
condition on $U$ above, to make $w$ periodic as well. However, the
structure of the action~\eqref{S5t} makes it clear that the ``vacuum
values'' $w(\c_n )$ of $w$ have to reproduce the charges on the
orbifold planes. We, therefore, define $w$ as
\begin{equation}
 w(\c ) = \int_0^\c d\tilde{\c}\sqrt{U(\tilde{\c})}
\end{equation}
which implies quasi-periodicity, that is,
\begin{equation}
 w(\c +v) = w(\c )+w(v)\; .
\end{equation}
We have plotted the typical form of $U$ and $w$ in Fig.~\ref{fig1}.

For much of our discussion the concrete form of the potential will
be irrelevant as long as the above conditions are met. A specific
example, however, is provided by the sine-Gordon potential
\begin{equation}
 U = m^2\left[1-\cos\left(\frac{2\p\c}{v}\right)\right]\; ,
\end{equation}
where $m$ is a constant. The associated superpotential $w$ is easily
obtained by integration.

\vspace{0.4cm}

This concludes the set-up of our model. Let us now discuss how,
precisely, this model corresponds to the brane-world
theory~\eqref{S5} introduced earlier. The simplest solution for $\c$
is to be in one of its vacuum states, that is, $\c=\c_n=nv$ for
some integer $n$, throughout space-time. In this case, the superpotential
and potential reduce to
\begin{equation}
 W = e^{-\F}w(\c_n)\; ,\qquad V = \frac{1}{3}e^{-2\F}w(\c_n)^2\; .
\end{equation}
Substituting this back into the action~\eqref{S5t} and comparing with
the M-theory result~\eqref{S5} shows that this precisely corresponds
to a situation without a bulk three-brane. In particular, one
concludes that the boundary charge $\a_1$ has to be identified with
the value $w(\c_n)=w(nv)=nw(v)$ of the superpotential at the
respective minimum~\footnote{Note that, in the absence of
three-branes, we have $\a_2=-\a_1$ from the cohomology
condition~\eqref{cohomology}.  Therefore, also the charge on the
second boundary is correctly being taken care of by our model.}. This
is, of course, the more precise reason why we have required the
superpotential to be quasi-periodic rather than periodic. Furthermore,
we learn that the elementary unit of charge $\s$ in the M-theory model
(see Eq.~\eqref{e0}) corresponds to $w(v)$, that is,
\begin{equation}
 \s = w(v) = \int_0^vd\x\sqrt{U(\c )}\; .\label{ident}
\end{equation}

The next more complicated solutions are kinks where the scalar field
$\c$ interpolates between two of its minima as one moves along
the orbifold direction. Due to the cross-couplings in the
action~\eqref{S5t} also the dilaton $\F$ and the metric
necessarily have a non-trivial profile in this case.
To find such solutions, an appropriate Ansatz is provided by
\begin{eqnarray}
 ds^2 &=& e^{2A(y)}dx^\m dx^\n\eta_{\m\n}+e^{2B(y)}dy^2 \label{A1}\\
 \F &=& \F (y) \\
 \c &=& \c (y)\; . \label{A3}
\end{eqnarray}
The four $y$-dependent functions $A$, $B$, $\F$, $\c$ are subject
to the second order bulk equations of motion to be derived
from the first line in~\eqref{S5t} and the boundary conditions
\begin{eqnarray}
e^{-B} A' &=& -\frac{1}{3}W = -\frac{1}{3}e^{-\F}w,\label{bc1}\\ 
e^{-B}\F ' &=& 2\frac{\partial W}{\partial\F} = -2e^{-\F}w, \label{bc2}\\
e^{-B}\c ' &=& e^{\F}\frac{\partial W}{\partial\c} = \frac{dw}{d\c}\; .
\label{bc3}
\end{eqnarray}
Here, the prime denotes the derivative with respect to $y$ and the
equations hold at both boundaries, that is, at $y=0$ and $y=\p\r$.
The first equality in each equation is easily derived from~\eqref{S5t}
including the boundary terms while the second one
follows from inserting the explicit form of the superpotential~\eqref{W}.

Instead of dealing with the second order equations to obtain explicit
solutions it is much simpler to consider the first order BPS-type
equations. Their existence is guaranteed by the special form of our
scalar field potential $V$ as being obtained from a
superpotential~\cite{Skenderis:1999mm}. Concretely, inserting the
Ansatz~\eqref{A1}--\eqref{A3} into the bulk part of the 
action~\eqref{S5t} one obtains an energy functional
\begin{eqnarray}
 E &\sim& \int dy e^{4A}\left[-6e^{-2B}{A'}^2+\frac{1}{4}e^{-2B}{\F '}^2
          +\frac{1}{2}e^{-\F-2B}{\c '}^2 + V\right], \nonumber\\
   &=&\int dy e^{4A}\left[\frac{1}{4}\left(e^{-B}\F '\mp 2\frac{\partial W}
      {\partial\F}\right)^2+\frac{1}{2}e^{-\F}\left(e^{-B}\c '\mp e^{\F}
      \frac{\partial W}{\partial\c}\right)^2-\frac{2}{3}\left(3e^{-B}A'
      \pm W\right)^2\right] \\
   &&\qquad\qquad\quad \pm\left[ e^{4\a}W\right]_{y=0}^{y=\p\r}\nonumber
\end{eqnarray}
which can be written in Bogomol'nyi perfect square form. This leads to the
following first order equations
\begin{eqnarray}
 e^{-B}A' &=& \mp\frac{1}{3}W = \mp\frac{1}{3}e^{-\F}w, \label{be1}\\
 e^{-B}\F ' &=& \pm 2\frac{\partial W}{\partial\F} = \mp 2e^{-\F}w,
   \label{be2}\\
 e^{-B}\c ' &=& \pm e^{\F}\frac{\partial W}{\partial\c}=\pm\frac{dw}{d\c}\; .
   \label{be3}
\end{eqnarray}
Again, the second equality in each line follows from inserting
the explicit superpotential~\eqref{W}. The scale factor $B$ is, of course,
a gauge degree of freedom and can, for example, be set to a constant.
It is clear then that equation~\eqref{be3} for $\c$ decouples
from the other two. This $\c$ equation is, in fact, exactly the
same first order equation one would derive for a single scalar field
$\c$ with potential $U$ in a flat background. It is, therefore, clear
and can be seen by direct integration, that this equation admits
kink solutions where $\c$ interpolates between a certain
minimum $\c= \c_n = nv$ of $U$ at $y\rightarrow -\infty$ and one of its
neighbouring minima at $y\rightarrow +\infty$. More precisely, for
the choice of the upper (lower) sign in Eq.~\eqref{be3} the
minimum at $\c = (n+1)v$ ($\c = (n-1)v$)
is approached for $y\rightarrow +\infty$. The corresponding
solutions for $A$ and $\F$ can then be obtained by inserting this
kink solution and integrating Eqs.~\eqref{be1} and \eqref{be2}.
In the next section, this will be carried out in a more precise way.
In addition, the solutions obtained in this way have to satisfy the
boundary conditions~\eqref{bc1}--\eqref{bc3}. Clearly, this is
automatically the case if the upper sign in the first order
equations~\eqref{be1}--\eqref{be3} has been chosen, that is, if
the kink interpolates between the minima $\c=nv$ and $\c=(n+1)v$
for increasing $y$. For the lower sign, on the other hand, there
is no chance to satisfy the boundary conditions and, hence, no
solutions of the type considered here exist in this case.
The interpretation of these results is straightforward. While
both types of kinks are on the same footing as far as the
bulk equations are concerned the boundary conditions distinguish
what should then be called an anti-kink, interpolating between $\c=nv$ and
$\c =(n-1)v$, from a kink, interpolating between $\c =nv$ and
$\c =(n+1)v$. While the latter represents a BPS solution of the
theory, the former carries the wrong orientation to be compatible
with the boundaries and, in fact, will only exist as a dynamical
object. This is in direct analogy with the properties of three-branes
and anti-three-branes in our original M-theory model~\eqref{S5}.

\vspace{0.4cm}

For the case of a kink, we would like to make this correspondence
with the M-theory model more precise. Let us consider a kink solution
to Eqs.~\eqref{be1}--\eqref{be3} and \eqref{bc1}--\eqref{bc3}
with the kink width being small (compared to the size of the orbifold)
and the core of the kink sufficiently away from the boundaries.
In this case, the profile for $\c$ and $w(\c )$ can be approximated
by a step-function. Specifically, we have $w(\c )\simeq n w(v)$ 
to the left of the kink and $w(\c )\simeq (n+1)w(v)$ to the right.
Inserting this into the equations~\eqref{be1}, \eqref{be2} and the 
boundary conditions~\eqref{bc1}, \eqref{bc2} for $A$ and $\F$
and solving the resulting system precisely leads to the BPS three-brane
solution given by Eqs.~\eqref{BPS1}, \eqref{BPS2}, \eqref{BPS5}.
The charges $\a_i$ appearing in this solution are given by
\begin{equation}
 \a_1= n\s\; ,\qquad \a_2= -(n+1)\s\; ,
 \qquad \a_3 = \s\; ,\label{ident1}
\end{equation}
where we have used our earlier identification~\eqref{ident} of
the superpotential value $w(v)$ with the elementary charge unit $\s$.
Hence, our model allows for a solution which can be interpreted
as a smooth version of the M-theory domain wall coupled to
a single-charged three-brane.

More generally, we would like to discuss the relation between
the action~\eqref{S5t} in the background of a kink
solution and the M-theory action~\eqref{S5}. To do this,
we should allow for fluctuations of the kink. It is
well-known~\cite{effdefect} that, for sufficiently small width,
the hypersurface prescribed by the kink's core is
a minimal surface and is, therefore, adequately described
by a Nambu-Goto action. Practically, this implies that the
kinetic term for $\c$ and the $U$ potential term in
the action~\eqref{S5t} can be effectively replaced by a Nambu-Goto
action describing the dynamics of the core of the kink. Of course,
this core has to be identified with the three-brane in the M-theory
model. It is easy to show that, by virtue of Eq.~\eqref{ident},
the tension in this effective Nambu-Goto action is given by $\s$
which is the correct value for a single-charged three-brane with
$\b_3 =1$. Further, the superpotential $w$ in such a kink background
can be effectively replaced by a step-function, as discussed
above. Using the identification~\eqref{ident1} of charges, it
is easy to see that the superpotential $w$ precisely equals 
the function $\a$ , defined in Eq.~\eqref{alpha}, in this limit.
As a consequence, the second potential term in \eqref{S5t}
proportional to $e^{-2\F}w^2$ precisely reproduces the bulk
potential in the M-theory action~\eqref{S5}. Similarly, the
boundary potentials in \eqref{S5t} match the boundary potentials
in \eqref{S5} using that $w(\c (y=0))\simeq n\s = \a_1$ and
$w(\c (y=\p\r ))\simeq (n+1)\s = -\a_2$. Although there are no
BPS anti-kink solutions, it is clear that a similar argument can
be made for the action~\eqref{S5t} in the background of an
anti-kink leading to the M-theory action~\eqref{S5} with an
anti-three-brane.

\vspace{0.4cm}

In summary, we have seen that the action~\eqref{S5t} in the
background of various vacuum configurations of the field $\c$
reproduces different versions of the M-theory effective
action~\eqref{S5}. For a constant field $\c$ located in one of the
minima of $U$, we have reproduced the M-theory action without three-branes.
For a kink (anti-kink) background with sufficiently small width
away from the boundaries we have obtaining
the M-theory action with a single-charged three-brane
(anti-three-brane). Note that, while from the viewpoint
of the smooth model~\eqref{S5t} these cases merely correspond
to different configurations of the field $\c$, they
represent different effective actions on the M-theory side.
As we have discussed, these different effective actions arise from
topologically distinct compactifications of the 11-dimensional
M-theory. While these compactifications are known to be related by
topology-changing transitions such as small-instanton transitions
these processes cannot be described by the action~\eqref{S5}.
What we have seen is, that our smooth defect model incorporates
a number of these topologically distinct configurations within
a single theory and, may, describe transitions between
them as the scalar field $\c$ evolves in time. In the subsequent sections,
we will study the simplest example for such a transition, namely
the collision of a kink with one of the boundaries.

\vspace{0.4cm}

A final comment concerns the question of multi-charged branes.
Clearly, multi-charged BPS three-branes with $\b_3>1$ are allowed in
the M-theory model~\eqref{S5}. However, our defect model~\eqref{S5t}
does not have exact BPS multi-kink solutions as long as the potential
$U$ is smooth at its minima. The reason is that, for smooth $U$, a
kink solution does not reach a minimum within a finite distance, as
can be easily seen from Eq.~\eqref{be3} with $U$ expanded around a
minimum. As a consequence, single-kink solutions cannot be ``stacked''
to produce exact multi-kink solutions. There are a number of options
available to remove this apparent discrepancy. Firstly, the
model~\eqref{S5t} as stands does have approximate multi-kink solutions
(with exponential accuracy) which could be identified with
multi-charged three-branes. Secondly, if the potential $U$ is
continuous but non-smooth at its minima a kink solution can reach a
minimum within a finite distance. There is no obstruction then to
build up exact multi-kinks by stacking single-kink solutions. Thirdly, some
multi-scalar field models are known to admit multi-kink
solutions~\cite{multikink}. So, we may generalise the action~\eqref{S5t} by
adding more than one scalar field. For the purpose of this paper,
we will not implement any of these possibilities explicitly but, rather,
focus on single-kink solutions in the following.


\section{The four-dimensional effective action of a kink solution}

We would now like to study one of the simplest dynamical processes
in the context of our defect model, namely the time-evolution
of a kink solution and its collision with a boundary. For a sufficiently
slow evolution this can be conveniently studied in the context of
the four-dimensional effective theory associated to~\eqref{S5t}
in the presence of a kink. The purpose of this section is to
compute this effective four-dimensional theory. As we will see,
this computation can be pushed a long way without specifying an
explicit potential $U$. We will, therefore, keep $U$ general throughout
this section. An explicit example for $U$ will be studied in the
next section.

\vspace{0.4cm}

Our first step is to write the kink solution in a form which
makes the dependence on the various integration constant (which
will be promoted to four-dimensional moduli fields later on) as
explicit as possible. We find that the kink solution to
\eqref{be1}--\eqref{be3} and \eqref{bc1}--\eqref{bc3} 
interpolating between the minima $\c = \c_n = nv$ and
$\c = \c_{n+1}=(n+1)v$ for increasing $y$ can be cast in the form
\begin{eqnarray}
 \c &=& C\left( e^\b\m^{-1}\left(y/\p\r -z\right)\right) ,
        \label{sol1} \\
 e^{\F} &=& e^{\f}\left( 1 + \e_0 e^{\b -\f}f(y,\b ,z)\right) ,\label{sol2}\\
 A &=& A_0+\frac{1}{6}\F, \label{sol3}\\
 B &=& \label{sol4}\b,
\end{eqnarray}
where we recall that $A$ and $B$ are the scale factors in the
five-dimensional metric as defined in Eq.~\eqref{A1}--\eqref{A3}. The
functions $C$ and $f$ in the above solution can be expressed in terms
of the potential as follows
\begin{eqnarray}
 C^{-1}(\c ) &=& \frac{1}{\p\r\m}\int_{(n+\frac{1}{2})v}^\c\frac{d\tilde{\c}}
               {\sqrt{U(\tilde{\c})}}\; , \qquad\label{C} \\
 f(y,\b ,z) &=& -\frac{2}{\p\r \e_0}\int_{y_0}^yd\tilde{y}\; w\left(C\left(
              e^\b\m^{-1}\left(\tilde{y}/\p\r -z\right)\right)
              \right)\; . \label{f}
\end{eqnarray}
Here $\f$, $\b$, $z$, $A_0$ and $y_0$ are integration constants, while
$\m$ is a constant which measures the width of the kink in units of
$\p\r$. It is clear from the form of the metric~\eqref{A1} that the
constant $A_0$ can be absorbed into the four-dimensional metric. As
we will see, it is, however, convenient to keep this constant
explicitly since it can be used to canonically normalise the
four-dimensional Einstein-Hilbert term. For our subsequent discussion,
let us define the average $\langle h \rangle$ of a function $h=h(y)$ over the
orbifold by
\begin{equation}
 \langle h \rangle = \frac{1}{\p\r}\int_0^{\p\r}dy\; h(y)\; .
\end{equation}
Since the constants $y_0$ and $\f$ really describe the same degree
of freedom, we can fix $y_0$ by requiring that $\langle f\rangle=0$. With this
convention, the integration constant $\f$ has a clear geometrical
interpretation, namely $e^\f$ represents the orbifold average of
the dilaton $e^\F$. Similarly, $e^\b$ measures the orbifold size in
units of $\p\r$. The final integration constant $z$ specifies the position
of the kink's core (the position where $\c=(n+\frac{1}{2})v$) in the
orbifold direction. Values $z\in [0,1]$ imply
that the kink's core is located within the boundaries of five-dimensional
space and is, hence, physically present. Further, $z\rightarrow 0,1$
indicates collision of the kink with one of the boundaries.
For $z\notin [0,1]$ the core is outside the physical region and we
can merely think of $z$ as the virtual position of the core were
space-time to continue beyond the boundaries. In this case,
the physical part of the kink, located between the boundaries, is
only its tail. In the limiting case $z\rightarrow\pm\infty$ the
kink disappears completely and we approach one of the trivial vacuum
states of the theory with either $\c = nv$ or $\c =(n+1)v$ throughout
five-dimensional space-time depending on whether
$z\rightarrow +\infty$ or $z\rightarrow -\infty$. Also note that the
function $C$, defined in Eq.~\eqref{C}, is independent of all
integration constants and can be computed for a given potential $U$.

\vspace{0.4cm}

We should now promote all integration constants in our kink
solution to four-dimensional moduli fields. This leads to three scalar
fields $(\f^I)=(\f ,\b ,z)$ and the four-dimensional effective metric
$g_{4\m\n}$. Accordingly, the Ansatz~\eqref{A1}--\eqref{A3} should
then be modified to
\begin{eqnarray}
 ds^2 &=& e^{2A(y,\f ^I)}dx^\m dx^\n g_{4\m\n} + e^{2B(y,\f ^I)}dy^2,
          \label{A1t} \\
 \F &=& \F (y,\f^I), \\
 \c &=& \c (y,\f^I)\; , \label{A3t}
\end{eqnarray}
where $A$, $B$, $\F$ and $\c$ are as in Eqs.~\eqref{sol1}--\eqref{sol4}
but with $(\f^I)=(\f ,\b ,z)$ now viewed as functions of the external
coordinates $x^\m$. 

We are now ready to compute the four-dimensional effective action. 
Inserting the Ansatz~\eqref{A1t}--\eqref{A3t} into the action~\eqref{S5t}
and integrating over the orbifold direction we obtain the
following result
\begin{equation}
 \tilde{S}_4 = -\frac{1}{2\k_P^2}\int_{M_4}\sqrt{-g_4}\left[
               \frac{1}{2}R_4+\frac{1}{2}G_{IJ}\partial_\m\f^I
               \partial^\m\f^J\right]\; . \label{S4t}
\end{equation}
The sigma-model metric $G_{IJ}$ is given by
\begin{equation}
 G_{IJ} = 2\left< e^{2A+B}\left[-3\partial_IA\partial_JA-3\partial_{(I}A
          \partial_{J)}B+\frac{1}{4}\partial_I\F\partial_J\F
          +\frac{1}{2}e^{-\F}\partial_I\c\partial_J\c \right]\right>,
 \label{G}
\end{equation}
where $\partial_I=\frac{\partial}{\partial\f^I}$ and $(\f^I)=(\f ,\b
,z)$. Further, in order to obtain an Einstein-frame action we have
required that
\begin{equation}
 \left< e^{2A+B}\right> = 1\; . \label{cannor}
\end{equation}
This indeed fixes the constant $A_0$ in Eq.~\eqref{sol3} to be
\begin{equation}
 e^{2A_0} = e^{-\b}\left< e^{\F /3}\right>^{-1}\; . \label{A0}
\end{equation}
The four-dimensional Planck scale $\k_P$ is defined by
\begin{equation}
 \k_P^2 = \frac{\k_5^2}{2\p\r}\; ,
\end{equation}
as usual. 

\vspace{0.4cm}

The remaining task is now to evaluate the expression~\eqref{G}
for the moduli-space metric using the kink
solution~\eqref{sol1}--\eqref{sol4}. This leads to fairly complicated
results, in general. There is, however, an approximation suggested
by the original M-theory model which simplifies matters considerably.
As discussed, the effective actions for heterotic M-theory
in Section 2 are valid only if the strong-coupling expansion parameter
\begin{equation}
 \e =\e_0 e^{\b - \f}  
\end{equation}
is smaller than one. We are, therefore, led to compute the moduli-space
metric~\eqref{G} in precisely this limit which corresponds to small
warping in the orbifold direction. Concretely, we will keep
terms up to ${\cal O}(\e )$ and neglect all terms of ${\cal O}(\e^2 )$
and higher in our computation. This implies a dramatic simplification
since the function $f$, which enters the kink solution Eq.~\eqref{sol2}
with an ${\cal O}(\e )$ suppression, drops out at this order.
Inserting~\eqref{sol2}--\eqref{sol4} and \eqref{A0} into~\eqref{G},
one then finds for the moduli-space metric
\begin{equation}
 G = \left(\begin{array}{ccc} \frac{1}{2}&0&0\\
     0&\frac{3}{2}+e^{-\f}\left<\left(\partial_\b\c\right)^2\right>&
     e^{-\f}\left<\partial_\b\c\partial_z\c\right>\\
     0&e^{-\f}\left<\partial_\b\c\partial_z\c\right>&
     e^{-\f}\left<\left(\partial_z\c\right)^2\right>
     \end{array}\right) + {\cal O}(\e^2 )\; .
\end{equation} 
Using the solution~\eqref{sol1} for $\c$ we finally obtain
\begin{eqnarray}
 G_{\f\f} &=& \frac{1}{2}, \label{Gff}\\
 G_{\b\b} &=& \frac{3}{2}+\left( e^{-\b}\m\right)^2\e_0 e^{\b-\f}
              \left[ J_2\left( e^\b\m^{-1}(1-z)\right)-J_2\left(
               -e^\b\m^{-1}z\right)\right], \label{Gbb}\\
 G_{\b z} &=& -e^{-\b}\m\e_0 e^{\b -\f}\left[J_1\left( e^\b\m^{-1}(1-z)
               \right)-J_1\left( -e^\b\m^{-1}z\right)\right], \label{Gbz}\\
 G_{zz}   &=& \e_0 e^{\b - \f}\left[ J_0\left( e^\b\m^{-1}(1-z)\right)
              -J_0\left( -e^\b\m^{-1}z\right)\right], \label{Gzz}
\end{eqnarray}
as the only non-vanishing components of $G$. Here, the functions $J_n$
are defined by
\begin{equation}
 J_n(x) = \frac{(\p\r )^2\m}{\e_0}\int_0^x d\tilde{x}\; \tilde{x}^n
          U(C(\tilde{x}))
        = \frac{1}{\s}\int_{C(0)}^{C(x)}d\c\left( C^{-1}(\c )\right)^n
          w' (\c )\; , \label{Jn}
\end{equation}
where we recall that the function $C$, defined in Eq.~\eqref{C}, can be
computed for any given potential $U$ and is, by itself, independent of
the moduli. The above result, good to ${\cal O}(\e )$, for the sigma model
metric explicitly displays the complete moduli dependence of $G$ and
its only implicit features are the dependence on the potential $U$ and
a simple integral thereof. We find it quite remarkable that the
calculation can be pushed this far without an explicit choice for
the potential $U$.       

\vspace{0.4cm}

The result~\eqref{Gff}--\eqref{Gzz} suggest the existence of another
expansion parameter besides $\e$, namely the quantity $e^{-\b}\m$.
It represent the ratio of the kink's width and the size of the orbifold.
Working in a thin-wall approximation where this ratio is much smaller
than one our results simplify even further. Clearly, we then have to
good accuracy
\begin{equation}
 G_{\b\b}=\frac{3}{2}\; ,\qquad G_{\b z} = 0\; .
\end{equation}
For the remaining non-trivial component $G_{zz}$ we can explicitly
carry out the integral \eqref{Jn} and find by inserting into
Eq.~\eqref{Gzz}
\begin{equation}
 G_{zz} = \e_0 e^{\b -\f}F(\b ,z)
\end{equation}
where
\begin{eqnarray}
 F(\b ,z) &=& \frac{1}{\s}\left[ w\left( C\left( e^\b\m^{-1}(1-z)\right)\right)
            - w\left( C\left( -e^\b\m^{-1}z\right)\right)\right], \label{F1}\\
          &=& \frac{1}{\s}\left[ w(y=\p\r )-w(y=0)\right] \; .\label{F2}
\end{eqnarray}
Here, the notation $w(y=0)$ ($w(y=\p\r $) indicates the value of
the superpotential evaluated for the kink solution
at the boundary $y=0$ ($y=\p\r$).
 
To summarise, in the limit of both the strong-coupling expansion parameter
and the ratio of wall to orbifold size being smaller than one, that
is,
\begin{equation}
 \e = \e_0 e^{\b -\f} < 1\; , \qquad \frac{\m}{e^\b} < 1\; ,
\end{equation}
the moduli-space metric for the kink solution is well-approximated by
\begin{equation}
 G = {\rm diag}\left(\frac{1}{2},\frac{3}{2},\e_0 e^{\b -\f}F(\b ,z)\right)
 \label{Gf}
\end{equation}
with associated four-dimensional effective action
\begin{equation}
 \tilde{S}_4 = -\frac{1}{2\k_P^2}\int_{M_4}\sqrt{-g_4}\left[
               \frac{1}{2}R_4+\frac{1}{4}\partial_\m\f\partial^\m\f
               +\frac{3}{4}\partial_\m\b\partial^\m\b
               +\frac{1}{2}\e_0 e^{\b -\f}F(\b ,z)\partial_\m z
               \partial^\m z\right]\; . \label{S4f}
\end{equation}
Here, the function $F$ is as defined in Eq.~\eqref{F1}.         

It is interesting to compare this four-dimensional effective action
to its counterpart~\eqref{S4} obtained in the M-theory case.
Obviously, the only difference arises in the kinetic term for $z$
where the function $F$ appears in~\eqref{S4f} but not in the
M-theory result~\eqref{S4}. A detailed comparison requires computing
this function from Eq.~\eqref{F1} by inserting an explicit potential
$U$. However, the qualitative features of $F$ can be easily read
off from the alternative expression~\eqref{F2}. It states that $F$
is the difference of the superpotential on the two boundaries
in units of $\s$ and, hence, it is simply
the "charge difference" between the two boundaries. Suppose, that
the kink's core is well within the physical space and away
from the boundaries, so that $z\in [0,1]$ and sufficiently different
from the boundary values $0$, $1$. The field $\c$ will then be
very close to the minimum $\c = nv$ at the $y=0$ boundary
and very close to the minimum $\c = (n+1)v$ at the
other boundary. The charge difference between the boundaries and, hence,
the function $F$, is, therefore, very close to one. If, on the
other hand, the virtual position of the kink's core is at $z>1$ ($z<0$) and
sufficiently away from the boundary, $\c$ will be close to the minimum 
$\c = nv$ ($\c = (n+1)v$) on {\em both} boundaries.
Hence the function $F$ is approximately zero in this case. This obviously
implies a non-trivial behaviour of $F$ close to the boundaries for
$z\simeq 0$ and $z\simeq 1$. As a result, for the kink being
inside the physical space and away from the boundaries by a distance
large compared to its width the effective action~\eqref{S4f}
completely agrees~\footnote{We recall that our kink carries a single
charge and we should, therefore, set $\b_3=1$ in Eq.~\eqref{q3} to obtain
perfect agreement.} with the M-theory result~\eqref{S4}. Conversely,
if the kink approaches one of the boundaries or collides with it,
that is, $z\rightarrow 0,1$, the function $F$ becomes non-trivial
and the effective theories~\eqref{S4f} and \eqref{S4} differ
substantially. It is clear then, that the effective theory~\eqref{S4f}
carries some memory of the presence of the boundaries while the
M-theory action~\eqref{S4} does not. For this reason,
studying the collision process in the context of~\eqref{S4f}
is an interesting problem which we will address in Section 6.


\section{An explicit example}

In this section, we consider the explicit example of the
double-well potential
\begin{equation}
 U = m^2(v^2-\c^2)^2\; ,  \label{Uex}
\end{equation}     
where $m$ is a constant. As stands this potential does, of course, not
satisfy our periodicity requirement for $U$. However, for our purposes
this is largely irrelevant since the single-kink solution in which we
are interested here probes the potential only between the two
minima~\footnote{One way to satisfy all earlier
requirements is to restrict the potential~\eqref{Uex} to the interval
$[-v,v]$ and continue it periodically outside. The subsequent results
do not depend on whether one works with this periodic version of the
potential or simply with its original form \eqref{Uex}.}. The
associated superpotential is given by
\begin{equation}
 w = m\c \left( v^2-\frac{1}{3}\c^2\right)\; .
\end{equation}
Hence the elementary charge unit $\s$ and $\e_0$ take the form
\begin{equation}
 \s = w(v)-w(-v) = \frac{4}{3}mv^3\; ,\qquad
 \e_0 = \p\r\s = \frac{4}{3}\p\r mv^3\; .
\end{equation} 

\vspace{0.4cm}

The kink-solution for this potential is of the general
form~\eqref{sol1}--\eqref{sol4} with the functions $C$ and $f$
given by
\begin{equation}
 C(x) = v\tanh (x) \label{Cex}
\end{equation}
and
\begin{equation}
 f = \frac{1}{\e_0e^\b}\left[ c-\frac{1}{3}v^2\tanh^2\x - \frac{4}{3}
     \ln (\cosh \x )\right]\; ,\qquad \x = \frac{e^\b}{\m}\left(
     \frac{y}{\p\r}-z\right)\; , \label{fex}
\end{equation}
where the thickness $\m$ of the kink can be identified as
\begin{equation}
 \m = \frac{1}{mv\p\r}\; .
\end{equation} 
The constant $c$ in Eq.~\eqref{fex} has to be fixed so that $<f>=0$,
as discussed before. This leads to an expression involving
di-logarithms and we will not carry this out explicitly.

\begin{figure}\centering
\includegraphics[height=9cm,width=9cm, angle=0]{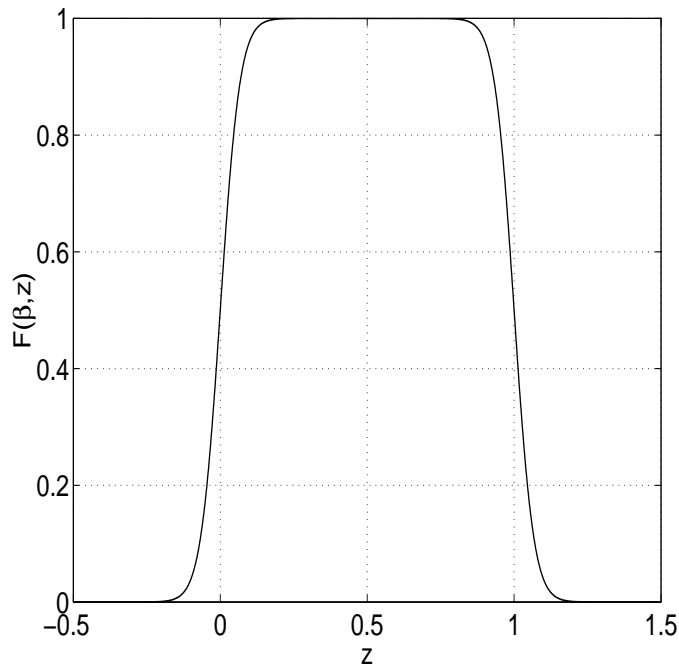}
        \caption{\emph{The function $F$ which enters the effective
         four-dimensional action of the kink as a function of $z$
         for $e^\b\m^{-1}=10$.}}
\label{fig2}
\end{figure}

Instead, we consider the limit where the strong-coupling expansion
parameter $\e$ remains small, so that $f$ becomes irrelevant and our general
result~\eqref{Gff}--\eqref{Gzz} holds. The functions $J_n$ can
now be explicitly computed inserting the potential~\eqref{Uex}
and~\eqref{Cex} into their definition~\eqref{Jn}. This leads to
\begin{equation}
 J_n(x) = \frac{3}{4}\int_0^xd\tilde{x}\frac{\tilde{x}^n}
          {\cosh^4\tilde{x}}\; .
\end{equation}
This, together with Eqs.~\eqref{Gff}--\eqref{Gzz} completely
determines the moduli-space metric for the double-well
potential as long as $\e <1$. While the above integrals can be
carried out for all relevant values $n=0,1,2$, the cases $n=1$
and $n=2$ lead to somewhat complicated expressions, the latter
involving a di-logarithm. However, $J_0$ takes the relatively simple form
\begin{equation}
 J_0(x) = \frac{1}{2}\tanh x + \frac{\sinh x}{4\cosh^3 x}\; .
 \label{J0}
\end{equation}
As is clear from the general case discussed in the previous section,
for a kink with small width, that is, $e^{-\b}\m < 1$, fortunately
$J_0$ is the only relevant function. In this limit, the moduli-space
metric is, therefore, given by the general form~\eqref{Gf} which we
repeat for convenience
\begin{equation}
 G = {\rm diag}\left(\frac{1}{2},\frac{3}{2},\e_0 e^{\b -\f}F(\b ,z)
     \right)\; .
 \label{Gex} 
\end{equation}
The function $F$, defined in Eq.~\eqref{F1}, now takes the explicit form
\begin{equation}
 F(\b ,z) = J_0\left( e^\b\m^{-1}(1-z)\right) - J_0\left( -e^\b\m^{-1}z
            \right)\; , \label{Fex}
\end{equation}   
where $J_0$ is given in Eq.~\eqref{J0}.
Inserting this result into~\eqref{S4f} completely determines the
four-dimensional kink effective theory for $\e < 1$ and $e^{-\b}\m
<1$. The function $F$ above indeed has the properties mentioned in the
previous section, namely $F\simeq 1$ for $z$ well inside the interval
$[0,1]$ and $F\rightarrow 0$ for $z\rightarrow\pm\infty$. The typical
shape of $F$ as a function of $z$ is shown in Fig.~\ref{fig2}.

\section{Kink evolution equations}

We will now study the time-evolution of the kink based on the effective
four-dimensional action derived in the previous section. The collision
of the kink with one of the boundaries will, of course, be of particular
interest.

We focus on simple time-dependent backgrounds and a metric
of Friedmann-Robertson-Walker form with flat spatial sections,
that is
\begin{eqnarray}
 ds_4^2 &=& -dt^2+e^{2\a (t)}d{\bf x}^2, \\
 \f^I &=& \f^I (t),
\end{eqnarray}
where $(\f^I)=(\f ,\b ,z)$. Let us first review the general
structure of the evolution equations for backgrounds of this form.
{}From the general sigma-model action~\eqref{S4t} one obtains the
equations of motion
\begin{eqnarray}
 3\dot{\a}^2 &=& \frac{1}{2}G_{IJ}\dot{\f}^I\dot{\f}^J, \\
 2\ddot{\a}+3\dot{\a}^2 &=& -\frac{1}{2}G_{IJ}\dot{\f}^I\dot{\f}^J, \\
 \ddot{\f}^I+3\dot{\a}\dot{\f}^I+\G_{JK}^I\dot{\f}^J\dot{\f}^K &=& 0\; ,
\end{eqnarray}
where $\G_{JK}^I$ is the Christoffel connection associated to the
sigma-model metric $G_{IJ}$ and the dot denotes the derivative with
respect to time. Adding the first two equations, we obtain an
equation for the scale factor $\a$ alone which can be immediately
integrated. Discarding trivial integration constants one finds
\begin{equation}
 \a = \frac{1}{3}\ln |t|\; .
\end{equation}
This power-law evolution with power $1/3$ is as expected
for a universe driven by kinetic energy only. We also remark that
we have, as usual, a $(-)$ branch, $t<0$, with decreasing $\a$ and
a future curvature singularity at $t=0$ and a $(+)$ branch, $t>0$,
with increasing $\a$ and a past curvature singularity at $t=0$.
Our subsequent results will apply to both branches although, for the
concrete discussion, we will mostly focus on the positive-time branch,
where the universe expands. We find it convenient to use the scale factor
$\a$, rather than $t$, as the time parameter in the following. The
remaining evolution equations can then be written in the form
\begin{eqnarray}
 {\f^I}^{\prime\prime}+\G_{JK}^I{\f^J}'{\f^K}' &=& 0, \label{geo} \\
 \frac{1}{2}G_{IJ}{\f^I}'{\f^J}' &=& 3\; ,\label{cons}
\end{eqnarray}
where the prime denotes the derivative with respect to $\a$. Hence,
the scalar fields $\f^I$, viewed as functions of the scale factor $\a$,
move along geodesics in moduli space, with initial conditions subject
to the constraint~\eqref{cons}.  
 
\vspace{0.4cm}

\begin{figure}\centering
\includegraphics[height=9cm,width=12cm, angle=0]{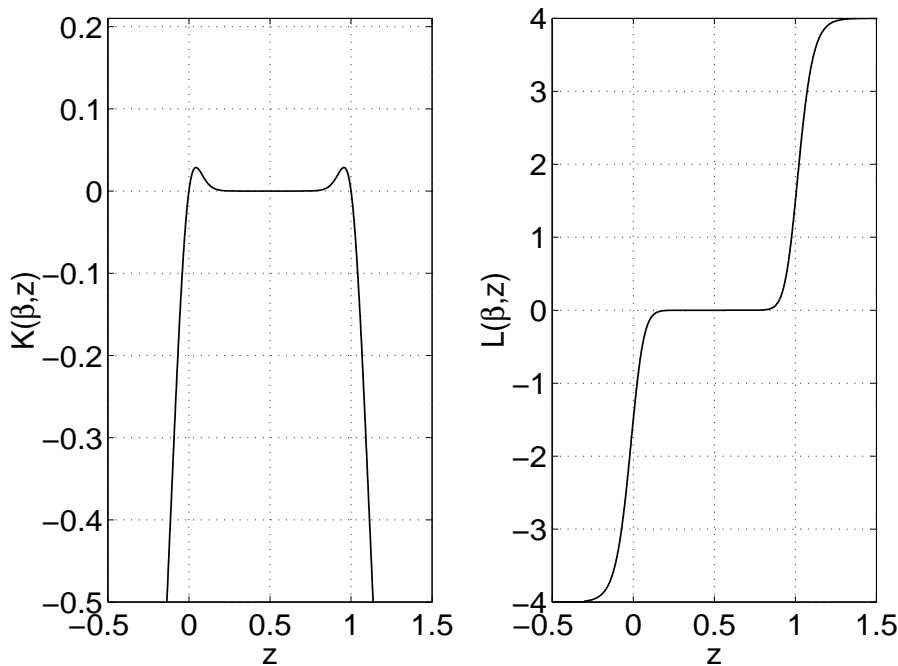}
        \caption{\emph{The function $K$ and $L$ 
         which enter the effective equations of motion for the kink
         as a function of $z$ for $e^\b\m^{-1}=10$.}}
\label{fig3}
\end{figure}

Let us now apply these equations to the moduli space metric for
the kink in a double-well potential, as computed in the previous
section. To keep the formalism as simple as possible we will focus
on the case of a small kink width, that is, $e^{-\b}\m\ll 1$. The moduli-space
metric is then specified by Eqs.~\eqref{Gex}, \eqref{Fex} and \eqref{J0}.
Inserting this metric into Eq.~\eqref{geo} we find
\begin{eqnarray}
 \f^{\prime\prime} +\e_0e^{\b -\f}F {z'}^2 &=& 0 ,\label{eom1}\\
 \b^{\prime\prime}-\frac{1}{3}\e_0e^{\b -\f}F\left( 1+e^\b\m^{-1}K\right)
  {z'}^2 &=& 0 ,\label{eom2}\\
 z^{\prime\prime} +(\b ' -\f ')z'+e^\b\m^{-1}K\b 'z'-\frac{1}{2}e^\b
  \m^{-1}L{z'}^2 &=& 0 ,\label{eom3}
\end{eqnarray}
while the constraint~\eqref{cons} turns into
\begin{equation}
 \frac{1}{4}{\f '}^2+\frac{3}{4}{\b '}^2+\frac{1}{2}\e_0 e^{\b -\f}F{z'}^2
    = 3\; . \label{eom4}
\end{equation}
The functions $K=K(\b ,z)$ and $L=L(\b ,z)$ are related to derivatives of
$F=F(\b ,z)$ and can be defined in terms of $J_0$, Eq.~\eqref{J0}, as
follows
\begin{eqnarray}
 F(\b ,z) &=& J_0\left( e^\b\m^{-1}(1-z)\right) - J_0\left( -e^\b\m^{-1}z
 \right) ,\\
 K(\b, z) &=& \frac{(1-z)J_0'\left( e^\b\m^{-1}(1-z)\right)
              +zJ_0'\left( -e^\b\m^{-1}z\right)}{F(\b ,z)}, \\
 L(\b ,z) &=& \frac{J_0'\left( e^\b\m^{-1}(1-z)\right)
              -J_0'\left( -e^\b\m^{-1}z\right)}{F(\b ,z)}\; .  
\end{eqnarray}
The typical shape of $F$ has been indicated in Fig.~\ref{fig2}.
Fig.~\ref{fig3} shows the shape of $K$ and $L$ as a function of $z$.

The equations of motion are generally quite complicated due to
these functions. However, as the figures show $F$, $K$ and $L$ are
non-trivial only in small regions around the boundaries with size
set by $\m e^{-\b}$ (the width of the kink relative to the orbifold size)
while they are relatively simple outside these critical regions.
It is, therefore, useful to discuss the asymptotic form of the equations
of motion away from the boundaries. First of all, for $z\in [0,1]$
and away from the boundaries we have
\begin{equation}
 F\simeq 1\; ,\qquad K\simeq 0\; ,\qquad L\simeq 0\; .
\end{equation}
Hence, for the kink being well inside the physical space the equations
of motion~\eqref{eom1}--\eqref{eom4} greatly simplify and become, in
fact, identical to the analogous equations derived from the M-theory
action~\eqref{S4}. 

On the other hand, for $z<0$ and away from the boundary we have
\begin{equation}
 F\simeq 0\; ,\qquad K\simeq 4z\; ,\qquad L \simeq -4\; ,
\end{equation}
There are analogous results for $z>1$ but we will focus on the case
$z<0$ for concreteness. Inserting these asymptotic expressions, we
see that the equations~\eqref{eom1}, \eqref{eom2} and
\eqref{eom4} for $\f$ and $\b$ decouple from the $z$ equation.
They become, in fact, the equations for freely rolling radii and can
be easily integrated to give
\begin{equation}
 \f = 3p_\f\a + \f_0\; ,\qquad \b = 3p_\b\a + \b_0 \label{rr}
\end{equation}
where $\f_0$ and $\b_0$ are arbitrary constants and the expansion
powers $p_\f$ and $p_\b$ satisfy the constraint
\begin{equation}
 p_\f^2+3p_\b^2 = \frac{4}{3}\label{crr}
\end{equation}
which follows from~\eqref{eom4}. The evolution of the kink can now be
studied in the background of these freely rolling radii. Inserting 
the above solutions for $\f$ and $\b$ into the equation for $z$,
Eq.~\eqref{eom3}, we find
\begin{equation}
 z^{\prime\prime}+3\d z^\prime+2\m_0^{-1}e^{3p_\b\a}(6p_\b z+z')z' = 0\; ,
 \label{zeq}
\end{equation}
where
\begin{equation}
 \m_0 = \frac{\m}{e^{\b_0}}
\end{equation}
is the width of the kink relative to the orbifold size initially at
$\a =0$ and
\begin{equation}
 \d = p_\b - p_\f\; .
\end{equation}
Hence, for $z<0$ and away from the boundary the evolution
of the kink is described by the single differential equation~\eqref{zeq}.     


\section{Kink dynamics and kink-boundary collision}

We should now study the solutions to the
system~\eqref{eom1}--\eqref{eom4}.  Given that our main interest is in
the collision of the kink with a boundary, ideally, we would like to
find solutions with $z\in [0,1]$ initially which evolve towards
$z\rightarrow 0$. Given the complexity of the equations, we cannot
possibly hope to achieve this analytically. Later, we will address
this problem numerically. However, some progress can be made
analytically as long as $z$ is away from the boundaries by using the
approximate equations for $ z\in [0,1]$ or $z<0$ discussed in the
previous section. One may hope that finding such analytical solutions
for the evolution up to shortly before and after the collision will
lead to a correct qualitative picture of the collision process,
roughly by gluing together these two types of solutions across the
critical boundary region. As we will see in our numerical analysis,
this is indeed the case.

\vspace{0.4cm} 

Let us start by looking at the case $z\in [0,1]$. As discussed above,
as long as $z$ is not too close to one of the boundaries, 
the equations of motion reduce to the ones obtained from the
M-theory effective action~\eqref{S4}. Their solutions have been found
in Ref.~\cite{Copeland:2001zp} and are explicitly given by
\begin{eqnarray}
 \f &=& 3p_{\f ,i}\a +3(p_{\f ,f}-p_{\f ,i})\ln\left( 1+e^{-3\d_i\a}
        \right)^{-\frac{1}{3\d_i}}+\f_0, \label{solf}\\
 \b &=& 3p_{\b ,i}\a +3(p_{\b ,f}-p_{\b ,i})\ln\left( 1+e^{-3\d_i\a}
        \right)^{-\frac{1}{3\d_i}}+\b_0, \label{solb}\\
 z &=& \frac{d}{1+e^{3\d_i\a}}+z_0\; .\label{solz}
\end{eqnarray}
Asymptotically, for $\a\rightarrow\pm\infty$, these solutions approach
freely rolling radii solutions for $\f$ and $\b$ while $z$ becomes constant.
The early (late) rolling radii solution is characterised by the
expansion powers $p_{\f ,i}$ and $p_{\b ,i}$ ($p_{\f ,f}$ and $p_{\b ,f}$).
Both sets of expansion powers are subject to the constraint
\begin{equation}
 p_{\f ,n}^2+3p_{\b ,n}^2 = \frac{4}{3}
\end{equation}
where $n=i,f$ and are related by the linear map
\begin{equation}
 \left(\begin{array}{l}p_{\b ,f}\\p_{\f ,f}\end{array}\right)
  = P\left(\begin{array}{l}p_{\b ,i}\\p_{\f ,i}\end{array}\right)\; ,
  \qquad P = \frac{1}{3}\left(\begin{array}{rr}1&1\\3&-1\end{array}
  \right)\; .\label{map}
\end{equation}
Further, we have defined the quantity
\begin{equation}
 \d_i = p_{\b ,i}-p_{\f ,i}
\end{equation}
which can be restricted, without loss of generality, to
\begin{equation}
 \begin{array}{lll}
   \d_i > 0&\qquad&(-)\mbox{ branch},\\
   \d_i < 0&\qquad&(+)\mbox{ branch},
 \end{array}
\end{equation}
We remark that $\d_f$, the analogous quantity at late times, is given by
\begin{equation}
 \d_f \equiv p_{\b ,f}-p_{\f ,f} = -\d_i
\end{equation}
as follows from the map~\eqref{map}.
The remaining integration constants $\f_0$, $\b_0$, $z_0$ and $d$ are
subject to the restriction
\begin{equation}
 \f_0-\b_0 = \ln\left(\frac{2\e_0d^2}{3}\right)\; .
\end{equation} 
Note that $z_0$ specifies the initial position of $z$ which
moves by a finite coordinate distance $d$ to its final position $z_0+d$.
 
What is the relevance of these solutions in our context?
First, we remind the reader that the above solutions play a
double-role as exact solution to the M-theory effective
action~\eqref{S4} and approximate solutions to the kink effective
theory if $z\in [0,1]$ and away from the boundaries. In their
former role they present another indication that the effective
M-theory action~\eqref{S4}, as it stands, is not adequate to describe
the collision process since the boundary values $z=0,1$ are in no
way singled out. In other words, $z$, as described by these solutions,
passes through the boundary without being effected at all. For this
reason, they will also be very useful for comparison with solutions
to the kink evolution equations, to explicitly see the boundary effect
in the latter. In their role as approximate solutions to the kink
evolution equations for $z\in [0,1]$ they tell us that the collision can
be arranged or avoided depending on a choice of initial conditions.
Indeed, the initial position $z_0$ of the kink and the coordinate
distance $d$ by which it moves can be chosen arbitrarily. Hence,
for the choice $z_0\in [0,1]$ and $z_0+d\in [0,1]$ (and both values
away from the boundaries) the entire evolution of the kink is
described by the solutions above and a collision with the boundary
never occurs. There is, however, a caveat to this argument. While
the kink becomes static asymptotically also the
strong-coupling expansion parameter $\e$ necessarily
diverges~\cite{Copeland:2001zp},
as can be seen from the above solutions. Therefore, we eventually
loose control of our approximation and the effective theory breaks down.
Clearly, from the arguments so far, we cannot guarantee that the kink
remains static when this happens. In this paper, we will not attempt
to improve on this, for example by going back to the five-dimensional
theory. Instead, we will be content with arranging a certain
characteristic behaviour, such as the kink becoming static,
to occur for some intermediate period of time before we loose
control over the effective theory.

\vspace{0.4cm}

Let us now analyse the evolution of the kink for $z<0$ and away from
the boundary (the case $z>1$ is similar, of course). In this case,
the system is adequately described by the single approximate
equation~\eqref{zeq} for $z$ while $\f$ and $\b$ are decoupled and
evolve according to one of the rolling radii solutions~\eqref{rr}.
Unfortunately, we did not succeed in integrating the $z$ equation
in general. However, we can find a number of partial solutions
which, as we will see, provide a good indication of the various,
qualitatively different types of $z$ evolution.

Let us consider the evolution of $z$ in the background of a special
rolling radii solution with a static orbifold, that is,
\begin{equation}
 p_\b = 0\; ,\qquad p_\f = \pm\frac{2}{\sqrt{3}}
\end{equation}
where the two possible values of $p_\f$ follow from Eq.~\eqref{crr}.
The equation~\eqref{zeq} for $z$ then simplifies to
\begin{equation}
 z^{\prime\prime}-3p_\f z'+2\m_0^{-1}{z'}^2 = 0\; .\label{zeq1}
\end{equation}
The general solution to this equation can be readily found to be
\begin{equation}
 z = z_0+\frac{v_c}{3p_\f}\ln\left[ 1+\frac{v_0}{v_c}\left(e^{3p_\f\a}
     -1\right)\right]\; , \label{zsol}
\end{equation}
where $z_0$ and $v_0$ are integration constants specifying the initial
position and velocity of $z$ at $\a =0$, that is, $z_0=z(\a =0)$ and
$v_0=z'(\a =0)$. Here, we are interested in solutions where $z_0$ is
negative and as close to the boundary as is compatible with the validity
of~\eqref{zeq1}. In addition, we need $v_0<0$ so $z$ evolves into
the region well-approximated by~\eqref{zeq1}. The parameter $v_c$ is
defined as
\begin{equation}
 v_c=\frac{3}{2}p_\f\m_0\; .
\end{equation}
Let us discuss the properties of this solution for an expanding
universe starting with the case $p_\f = +2/\sqrt{3}$. It is easy to
see from Eq.~\eqref{zsol} that, independent of the initial velocity $v_0$,
$z$ always diverges to $-\infty$ at some finite value of the scale
factor $\a$, in this case. For $p_\f = -2/\sqrt{3}$, however,
the situation is somewhat more complicated and depends on the relation between
$|v_0|$ and $|v_c|$. One has to distinguish the three cases
\begin{itemize}
 \item[1)] $|v_0|<|v_c|$~: $z$ converges exponentially to a constant
 \item[2)] $|v_0|=|v_c|$~: $z$ diverges to $-\infty$ as $\a\rightarrow\infty$
 \item[3)] $|v_0|>|v_c|$~: $z$ diverges to $-\infty$ at a finite value
                           of $\a$.
\end{itemize}
Hence, we see that $v_c$ plays the role of a critical velocity. As we
will confirm later, these three cases already represent the three
types of qualitatively different behaviour which can be observed
for the full $z$-equation~\eqref{zeq} or even the complete
system~\eqref{eom1}--\eqref{eom4}.

We should remark, though, that
the second case $|v_0|=|v_c|$ while typical in that $z$ diverges as
$\a\rightarrow\infty$ is not representative as far as the nature of
the divergence is concerned. While its divergence is linear in $\a$,
the more characteristic case is an exponential divergence in $\a$.
The existence of such exponential divergences can be seen from the
special solution
\begin{equation}
 z = \frac{\m_0p_\f}{2p_\b}e^{-3p_\b\a}
\end{equation}
to Eqs.~\eqref{zeq}. While this represents an exact solution for all
values of $p_\f$ and $p_\b$ we have to restrict signs to
$p_\b <0$ and $p_\f >0$ so that $z$ is negative and moves
towards $-\infty$. Within this range of $p_\f$ and $p_\b$, however,
the above solution shows an exponential divergence of $z$ as
$\a\rightarrow\infty$. 

\vspace{0.4cm}

After having identified the qualitatively different types of $z$
evolution we can now ask more systematically, based on the $z$
equation~\eqref{zeq}, which type is realized for a given set of
parameters and initial conditions. As can be seen from a rescaling
of $z$ in Eq.~\eqref{zeq} the type of evolution cannot depend on the
value of $\m_0$. The only possible dependence is, therefore, on
$p_\b$ (recall that, for given $p_\b$, $p_\f$ is determined,
up to a sign, from Eq.~\eqref{crr}) and the initial velocity
$v_0=z'(\a =0)$. A relevant question in this context concerns
the stability of the solution $z=$ const which can be viewed as
the limit of the exponentially converging case 1. Writing
\begin{equation}
 z=z_0+\z (\a )\; ,
\end{equation}
where $z_0<0$, the linearised evolution equation for $\z$ is, from
Eq.~\eqref{zeq}, given by
\begin{equation}
 \z^{\prime\prime} = -3\left[\d +4\m_0^{-1}z_0p_\b e^{3p_\b\a}\right]\z '\; .
\end{equation}
We conclude that the solution $z=$ const can only be stable if
\begin{equation}
 p_\b < 0\; \quad\mbox{and}\quad \d = p_\b -p_\f > 0\; .\label{stablecond}
\end{equation}
It is only then that we expect the first case of convergent $z$ to
be realized.

This can indeed by verified by a numerical integration of Eq.~\eqref{zeq}.
Solutions with converging $z$ exist if and only if the
conditions~\eqref{stablecond} are satisfied and, in addition, if the
initial velocity $|v_0|$ is smaller than a certain critical velocity
$v_c$. A simple scaling argument shows that
\begin{equation}
 v_c = h(p_\b ,p_\f )\m_0
\end{equation}
where $h$ is a function which, from the numerical results, turns out to
be of ${\cal O}(1)$ and slowly varying. What happens outside the
region~\eqref{stablecond}? If we leave this range by crossing $p_\b =0$
we find for small positive $p_\b$ and $|v_0|$ below the critical velocity
that $z$ still converges at first but then, in accordance with our
analytic argument, develops an instability, which drives it to $-\infty$
at finite $\a$. The intermediate stable phase gradually disappears as
one increases $p_\b$. For $p_\b >0$ and $|v_0|$ above the critical velocity
one always finds divergence to $-\infty$ at finite $\a$. Hence, for
$p_\b >0$ we are always in the third case above. 
As we leave the region~\eqref{stablecond} crossing $\d =0$ we find
case 2 is realized below and case 3 above the critical velocity.
However, as $\d$ becomes more negative, the critical velocity
decreases rapidly until we are left with case 3 only.

In summary, the converging case 1 is only found in the
range~\eqref{stablecond} and for initial velocities smaller than
a certain critical value while otherwise $z$ always diverges to
$-\infty$ typically according to case 3 at finite scale factor $\a$.

\vspace{0.4cm}

We can now try to combine the information we have gathered about the
evolution of the system before and after the collision to set up
criteria which will allow us to decide the outcome of a collision
process. Let us consider a particular solution~\eqref{solf}--\eqref{solz}
for the evolution inside the interval $z\in [0,1]$. As we have already
mentioned, the distance by which the kink moves is a free parameter
so a collision may never occur. Then, this solution describes
the full evolution of the system as far as it is accessible within
the four-dimensional effective theory. On the other hand, if initial
conditions are chosen so that a collision does occur, the particular
solution~\eqref{solf}--\eqref{solz} will determine the velocities
$z'_{\rm col}$, $\f '_{col}$ and $\b '_{col}$ right
before the collision. We can then, approximately, identify
$v_0\simeq z'_{\rm col}$, $3 p_\f\simeq \f '_{col}$
and $3 p_\b\simeq \b '_{col}$ and apply the previous results for
the evolution at $z<0$. One concludes that
only for a very low-impact collision with small $z'_{\rm col}$
and an orbifold size which, at collision, decreases less rapidly
than the dilaton, that is $\b '_{col}<0$ and
$\f '_{col}-\b '_{col}>0$, does $z$ converge to a constant.
Otherwise $z$ diverges to $-\infty$ and this can, in fact, be viewed
as the generic case.

\vspace{0.4cm}

Of course, the criteria above may be somewhat inaccurate since we
have ignored the complicated structure of the evolution equations
near the boundary. We have, therefore, numerically integrated the
full system~\eqref{eom1}--\eqref{eom4} to test the above criteria
for the outcome of a collision process. It turns out that, in broad
terms, the picture remains qualitatively the same.

 Starting with $z$ near zero inside the $[0,1]$ interval, we went
around the ellipse $(\f '_{\rm col})^2+3 (\b '_{\rm col})^2\simeq 12$.
Note that in this case the exact identity cannot be observed since the
constraint equation Eq.~\eqref{eom4} includes an extra term
proportional to $(z'_{\rm col})^2$. Nevertheless the correction is
always small since we set $\f _{\rm col}-\b_{\rm col}$ to a large
negative value. This makes the initial value for $\epsilon$ very small
and allows us, for the cases where $\e$ grows, to follow the evolution
for longer times until $\epsilon\simeq 1$ and the four-dimensional
effective theory breaks down. We also chose a large initial $\b_{\rm
col}$ so that $e^{-\b}\mu$ remains as small as possible during the
evolution, for the cases with $\b'_{\rm col}<0$. In all cases we set
$\epsilon_0=1$ and $\mu=0.2$.

For each of these sets of initial conditions we then varied
$z'_{\rm col}$ from zero upwards and looked for changes in the large
time behaviour of $z$. The numerical results were obtained by evolving
Eqs.~\eqref{eom1}--\eqref{eom3} using a fourth-order fixed step 
Runge-Kutta method. The accuracy of the method was checked by
confirming that the constraint equation Eq.~\eqref{eom4} was satisfied
throughout the evolution.  The individual terms on the left hand side of 
Eq.~\eqref{eom4} should sum to 3, and typically after 2000 
time-steps of size 0.01 the deviation from this value was smaller 
than 0.01\%.  In the worst cases, where the equations of motion are 
no longer valid because one of the assumptions have broken down, 
the sum never gets above 0.2\%
     
\begin{figure}\centering
\includegraphics[height=9cm,width=9cm, angle=0]{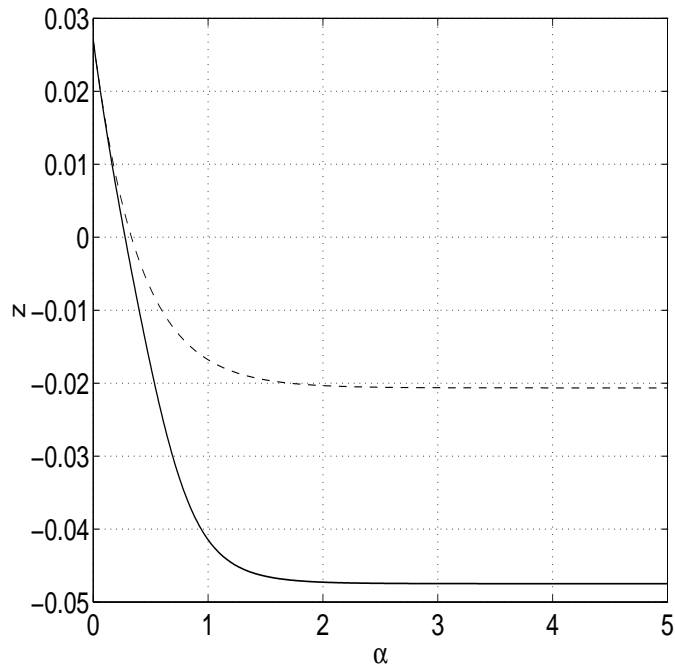}
        \caption{\emph{Position modulus $z$ for the kink
                 (solid line) and M-theory three-brane (dashed line)
                 as a function of the scale factor $\a$. The
                 initial conditions have been chosen as
                 $z_{\rm col}=0.027$, $z'_{\rm col}=-0.12$,
                 $\b_{\rm col}=2.0$, $\b'_{\rm col}=-0.72$,
                 $\f_{\rm col}=16.14$, $\f'_{\rm col}=-3.23$ }}
\label{fig4}
\end{figure}

 In Fig.~\ref{fig4} we have an example of the first type of behaviour,
for a small negative value of $\b'_{\rm col}$. After
crossing the boundary at $z=0$ the kink relaxes to a stable
constant solution. For early times this solution matches the one
obtained from the M-theory effective action for the same initial
conditions. Nevertheless, as soon as the kink approaches the boundary
the two start differing, converging to different asymptotic 
values.
     
\begin{figure}\centering
\includegraphics[height=9cm,width=9cm, angle=0]{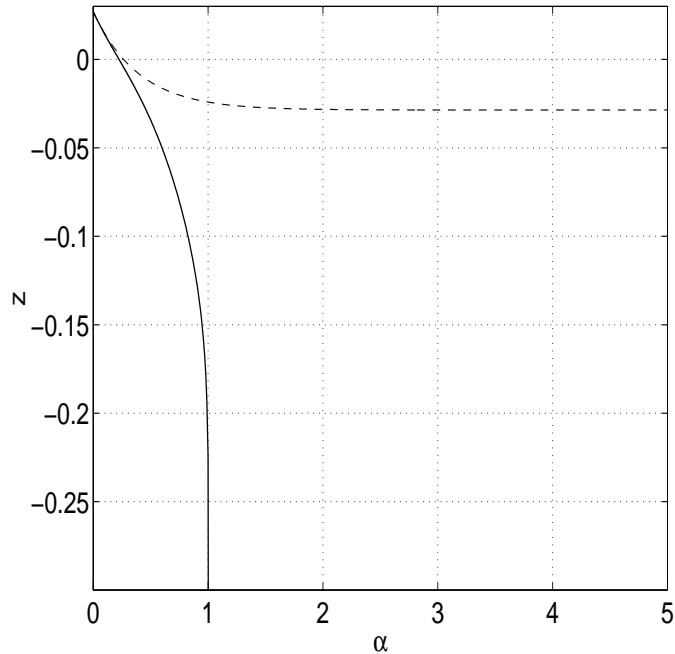}
        \caption{\emph{Same as in Fig.~\ref{fig4} but with
                       $z'_{\rm col}=-0.14$.}}
\label{fig5}
\end{figure}

For a slightly higher value of the initial velocity the difference is
even more striking, as shown in Fig.~\ref{fig5}. In this case $z$
diverges in finite time, indicating that we are above the critical
velocity. This third case turns out to be the most common, as already
observed in the simplified system. Only for $\b'_{\rm col}<0$ and
$\b'_{\rm col}-\f'_{\rm col}>0$ and $z'_{\rm col}$ below the critical
velocity does the system avoid singular behaviour.
     
\begin{figure}\centering
\includegraphics[height=9cm,width=9cm, angle=0]{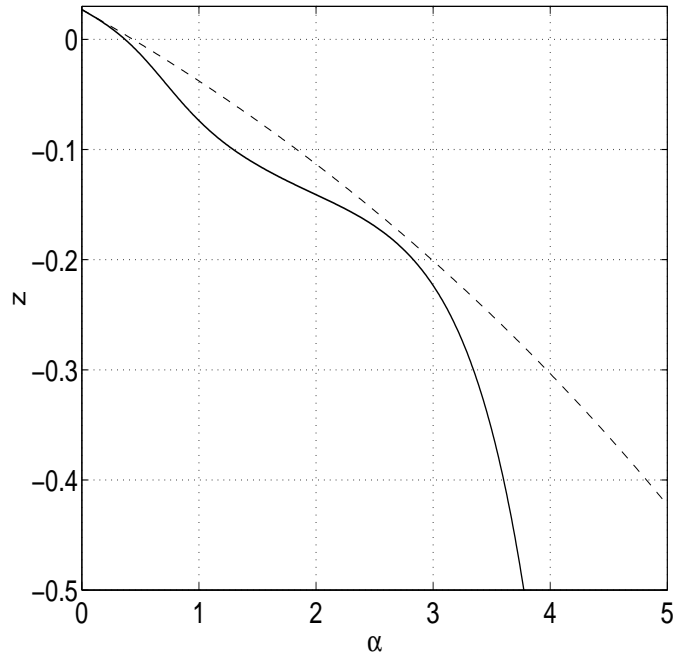}
        \caption{\emph{Position modulus $z$ for the kink
                 (solid line) and M-theory three-brane
                 (dashed line) as a function of the scale factor $\a$.
                 The initial conditions have been
                 chosen as $z_{\rm col}=0.027$, $z'_{\rm col}=-0.060$,
                 $\b_{\rm col}=2.0$, $\b'_{\rm col}=-1.77$,
                 $\f_{\rm col}=14.75$, $\f'_{\rm col}=-1.61$}}
\label{fig6}
\end{figure}

 In Fig.~\ref{fig6} we have an example for a solution
corresponding to case 2. Here both $\b'_{\rm col}$ and $\f'_{\rm col}$
are negative and we are below the critical velocity. As a consequence
of $\d \simeq \b'_{\rm col}-\f'_{\rm col}<0$, $z$ does not relax to a
constant but its magnitude increases exponentially instead. In this case 
the solution has to be taken with care, since $e^{-\b}\mu$
quickly becomes large in the exponential regime and the equations of
motion stop providing a reliable approximation.

Finally we have checked that once we go above the critical velocity,
$z$ always diverges for finite $\alpha$. It is well known that in
$\phi^4$ theory when a kink anti-kink collision takes place, above a 
certain limit velocity, they reflect and bounce back 
\cite{Campbell} (for lower velocities they can either reflect or
form a bound state). This behaviour is a consequence of a resonance 
effect between the kink pair and higher field modes, so we should not be 
surprised not to observe it in the context of our four-dimensional
effective action. This does not yet exclude the possibility of a
bounce in a high-velocity regime which is accessible only in the
context of the full five dimensional theory, a question which is
currently under investigation~\cite{inprep}.

\vspace{0.4cm}

What do these results imply in terms of the five-dimensional
defect model~\eqref{S5t}? As we have seen, if $z$ starts
its evolution within the interval $[0,1]$ and subsequently collides
with a boundary (at $z=0$) it is generically driven to $-\infty$
very rapidly. It should be stressed that the $z$ kinetic energy
remains finite at this singularity. Nevertheless, we do expect
the effective four-dimensional theory to break down eventually,
as $z\rightarrow -\infty$. This is because some of the higher-order
term we have neglected are likely to grow with $z$, in a way similar
to the linear $z$ term in Eq.~\eqref{zeq}. However, at least
for sufficiently small expansion parameters $\e$ and $\m e^{-\b}$
the four-dimensional theory will be valid some way into the singularity.
Hence, we can conclude that a five-dimensional kink,
interpolating between the vacua $\c = nv$ and $\c = (n+1)v$ which
collides with the boundary at $z=0$ effectively disappears and
leaves the field $\c$ in the vacuum state $\c = (n+1)v$ (and
an analogous statement holds for collision with the boundary at $z=1$).
{}From the M-theory perspective, such a process corresponds
to a transition
\begin{equation}
 (\b_1,\b_2,\b_3) = (n,-(n+1),1)\longrightarrow
 (\b_1,\b_2,\b_3) = (n+1,-(n+1),0) 
\end{equation}
between two different sets of charges and, hence, topologically
different compactifications.
     

\section{Conclusion and outlook}

In this paper, we have presented a toy defect model for
five-dimensional heterotic brane-world theories, where three-branes
are modelled by kink solutions of a bulk scalar field $\c$. We have
shown that the vacuum states of this defect model correspond to a
class of topologically distinct M-theory models characterised by the
charges $\b_1$ and $\b_2$ on the boundaries and the three-brane charge
$\b_3$. Specifically, we have seen that a state where $\c$ equals one
of the minima $\c = \c_n=nv$ of the potential, where $n\in{\bf Z}$,
corresponds to a state with charges $(\b_1,\b_2,\b_3)=(n,-n,0)$, that
is, an M-theory model without three-branes. If, on the other hand,
$\c$ represents a kink solution interpolating between the minima $\c =
nv$ and $\c = (n+1)v$ the associated M-theory charges are
$(\b_1,\b_2,\b_3)=(n,-(n+1),1)$ corresponding to a model with a
single-charged three-brane. 

We have computed the effective four-dimensional action associated
to a kink solution and have studied the time-evolution of a kink in
this context. Our results show that, generically, a collision of
the kink with a boundary will lead to a transition between the
two types of vacua mentioned above. In other words, the kink
will disappear after collision which corresponds to a transition between
a state with a single-charged three brane and a state without a three-brane.

\vspace{0.4cm}

There are several interesting directions which may be pursued on the
basis of these results. Clearly, our original M-theory model as well
as the associated defect model are rather simple and a number of
possible extension and modifications come to mind. First of all, we
may try to modify our defect model by including more than one
additional bulk scalar field, in particular to allow for exact BPS
multi-kink solutions. One may ask whether the defect model can be
embedded into a five-dimensional $N=1$ supergravity theory as is the
case for the original M-theory model. Further, there are a number of
generalisations of five-dimensional heterotic M-theory, such as
including a more general set of moduli fields~\cite{Lukas:1998tt}, which one
may try to implement into the defect model. For example, including
the general set of Kahler moduli would allow one to study
topological transitions of the underlying Calabi-Yau space through
flops, in addition to the types of topology change considered in this paper.

\vspace{0.4cm}

Perhaps the most interesting direction is to study the evolution of
more complicated configurations of our defect model~\eqref{S5t}.
For example, one could envisage evolving the field $\c$ from some
initial (say thermal) distribution to see which type of brane-network
develops at late time~\cite{inprep}. In particular, one would like to
answer the important question whether the system can evolve from a
brane-gas to a brane-world state. If this is indeed what happens
such an approach will lead to predictions for the late-time brane-world
that has evolved, given a certain class of plausible initial states.
Concretely, within the context of the simple model presented in this
paper, we may expect predictions for the charges $\b_i$ in this case.
As we have discussed, the values of these charges are correlated with
important properties of the theory such as the type of gauge group.
Optimistically, we may therefore hope that our approach leads to
prediction for such important low-energy data, at least within a
restricted class of associated M-theory compactifications.


\vspace{1cm}

\noindent
{\Large\bf Acknowledgements}\\
A.~L.~is supported by a PPARC Advanced Fellowship.
N.~D.~A.~is supported by a PPARC Post-Doctoral Fellowship.
     


\begin{thebibliography}{99}

\bibitem{Horava:1996ma}
P.~Horava and E.~Witten,
``Eleven-Dimensional Supergravity on a Manifold with Boundary,''
Nucl.\ Phys.\ B {\bf 475} (1996) 94
[hep-th/9603142].

\bibitem{Witten:1996mz}
E.~Witten,
``Strong Coupling Expansion Of Calabi-Yau Compactification,''
Nucl.\ Phys.\ B {\bf 471} (1996) 135
[hep-th/9602070].

\bibitem{Horava:1996vs}
P.~Horava,
``Gluino condensation in strongly coupled heterotic string theory,''
Phys.\ Rev.\ D {\bf 54} (1996) 7561
[hep-th/9608019].

\bibitem{Lukas:1997fg}
A.~Lukas, B.~A.~Ovrut and D.~Waldram,
``On the four-dimensional effective action of strongly coupled heterotic  string theory,''
Nucl.\ Phys.\ B {\bf 532} (1998) 43
[hep-th/9710208].

\bibitem{Lukas:1998hk}
A.~Lukas, B.~A.~Ovrut and D.~Waldram,
``Non-standard embedding and five-branes in heterotic M-theory,''
Phys.\ Rev.\ D {\bf 59} (1999) 106005
[hep-th/9808101].

\bibitem{Lukas:1998yy}
A.~Lukas, B.~A.~Ovrut, K.~S.~Stelle and D.~Waldram,
``The universe as a domain wall,''
Phys.\ Rev.\ D {\bf 59} (1999) 086001
[hep-th/9803235].

\bibitem{Ellis:1998dh}
J.~R.~Ellis, Z.~Lalak, S.~Pokorski and W.~Pokorski,
``Five-dimensional aspects of M-theory dynamics and supersymmetry  breaking,''
Nucl.\ Phys.\ B {\bf 540} (1999) 149
[hep-ph/9805377].

\bibitem{Lukas:1998tt}
A.~Lukas, B.~A.~Ovrut, K.~S.~Stelle and D.~Waldram,
``Heterotic M-theory in five dimensions,''
Nucl.\ Phys.\ B {\bf 552} (1999) 246
[hep-th/9806051].

\bibitem{Andreas:1999ei}
B.~Andreas,
``On vector bundles and chiral matter in N = 1 heterotic
compactifications,''
JHEP {\bf 9901} (1999) 011
[hep-th/9802202].
     
\bibitem{Curio:1998vu}
G.~Curio,
``Chiral matter and transitions in heterotic string models,''
Phys.\ Lett.\ B {\bf 435} (1998) 39
[hep-th/9803224].

\bibitem{Donagi:1999xe}
R.~Donagi, A.~Lukas, B.~A.~Ovrut and D.~Waldram,
``Non-perturbative vacua and particle physics in M-theory,''
JHEP {\bf 9905} (1999) 018
[hep-th/9811168].

\bibitem{Donagi:1999gc}
R.~Donagi, A.~Lukas, B.~A.~Ovrut and D.~Waldram,
``Holomorphic vector bundles and non-perturbative vacua in M-theory,''
JHEP {\bf 9906} (1999) 034
[hep-th/9901009].

\bibitem{Donagi:1999ez}
R.~Donagi, B.~A.~Ovrut, T.~Pantev and D.~Waldram,
``Standard models from heterotic M-theory,''
hep-th/9912208.

\bibitem{Donagi:2000vs}
R.~Donagi, B.~A.~Ovrut, T.~Pantev and D.~Waldram,
``Non-perturbative vacua in heterotic M-theory,''
Class.\ Quant.\ Grav.\  {\bf 17} (2000) 1049.

\bibitem{Donagi:2000zs}
R.~Donagi, B.~A.~Ovrut, T.~Pantev and D.~Waldram,
``Standard-model bundles,''
math.ag/0008010.

\bibitem{Witten:1996gx}
E.~Witten,
``Small Instantons in String Theory,''
Nucl.\ Phys.\ B {\bf 460} (1996) 541
[hep-th/9511030].

\bibitem{Ganor:1996mu}
O.~J.~Ganor and A.~Hanany,
``Small $E_8$ Instantons and Tensionless Non-critical Strings,''
Nucl.\ Phys.\ B {\bf 474} (1996) 122
[hep-th/9602120].

\bibitem{Ovrut:2000qi}
B.~A.~Ovrut, T.~Pantev and J.~Park,
``Small instanton transitions in heterotic M-theory,''
JHEP {\bf 0005} (2000) 045
[hep-th/0001133].

\bibitem{DeWolfe:1999cp}
O.~DeWolfe, D.~Z.~Freedman, S.~S.~Gubser and A.~Karch,
``Modeling the fifth dimension with scalars and gravity,''
Phys.\ Rev.\ D {\bf 62} (2000) 046008
[hep-th/9909134].

\bibitem{Brandle:2001ts}
M.~Brandle and A.~Lukas,
``Five-branes in heterotic brane-world theories,''
Phys.\ Rev.\ D {\bf 65} (2002) 064024
[hep-th/0109173].

\bibitem{Derendinger:2001gy}
J.~Derendinger and R.~Sauser,
``A five-brane modulus in the effective N = 1 supergravity of M-theory,''
Nucl.\ Phys.\ B {\bf 598} (2001) 87
[hep-th/0009054].

\bibitem{Strominger:et}
A.~Strominger,
``Heterotic Solitons,''
Nucl.\ Phys.\ B {\bf 343} (1990) 167
[Erratum-ibid.\ B {\bf 353} (1991) 565].

\bibitem{Copeland:2001zp}
E.~J.~Copeland, J.~Gray and A.~Lukas,
``Moving five-branes in low-energy heterotic M-theory,''
Phys.\ Rev.\ D {\bf 64} (2001) 126003
[hep-th/0106285].

\bibitem{Skenderis:1999mm}
K.~Skenderis and P.~K.~Townsend,
``Gravitational stability and renormalization-group flow,''
Phys.\ Lett.\ B {\bf 468} (1999) 46
[hep-th/9909070].

\bibitem{effdefect} 
F.~Bonjour, C.~Charmousis and R.~Gregory,
``The dynamics of curved gravitating walls,''
Phys.\ Rev.\ D {\bf 62} (2000) 083504
[gr-qc/0002063];
%
B.~Carter and R.~Gregory,
``Curvature corrections to dynamics of domain walls,''
Phys.\ Rev.\ D {\bf 51} (1995) 5839
[hep-th/9410095].

\bibitem{multikink} 
%
M.~A.~Shifman,
``Degeneracy and continuous deformations of supersymmetric domain walls,''
Phys.\ Rev.\ D {\bf 57} (1998) 1258
[hep-th/9708060];
%
C.~Bachas, J.~Hoppe and B.~Pioline,
``Nahm equations, N = 1* domain walls, and D-strings in AdS(5) x S(5),''
JHEP {\bf 0107} (2001) 041
[hep-th/0007067];
%
J.~P.~Gauntlett, D.~Tong and P.~K.~Townsend,
``Multi-domain walls in massive supersymmetric sigma-models,''
Phys.\ Rev.\ D {\bf 64} (2001) 025010
[hep-th/0012178];
%
A.~A.~Izquierdo, M.~A.~Leon and J.~M.~Guilarte,
``The kink variety in systems of two coupled scalar fields in two  space-time dimensions,''
Phys.\ Rev.\ D {\bf 65} (2002) 085012
[hep-th/0201200];
%
D.~Tong,
``The moduli space of BPS domain walls,''
Phys.\ Rev.\ D {\bf 66} (2002) 025013
[hep-th/0202012].
     
\bibitem{Campbell}
D.~K.~Campbell, J.~F.~Schonfeld and C.~A.~Wingate
``Resonance structure in kink-antikink interactions in $\phi^4$ theory,''
Physica {\bf 9} D (1983) 1. 
     
\bibitem{inprep} N.~D.~Antunes, E.~J.~Copeland, M.~Hindmarsh and
A.~Lukas, in preparation.

\end{thebibliography}
\end{document}